\documentclass[twocolumn,fp,seceq]{jpsj3}
\usepackage{txfonts}
\usepackage{graphicx}
\usepackage{dcolumn}
\usepackage{bm}
\usepackage[usenames]{color} 
\usepackage{amssymb} 
\usepackage{amsmath} 
\usepackage[utf8]{inputenc} 
\usepackage{braket}
\usepackage{mathptmx}
\usepackage[usenames]{color}
\newcommand{\MyChi}[2]{\chi_{#1;#2}}

\newcommand{\Uell}{u_{\ell {\bm k}}}
\newcommand{\Uellp}{u_{\ell' {\bm k}}}
\newcommand{\Uellpp}{u_{\ell'' {\bm k}}}
\newcommand{\Eell}{\varepsilon_{\ell}}
\newcommand{\Eellp}{\varepsilon_{\ell '}}
\newcommand{\Eellpp}{\varepsilon_{\ell ''}}

\newcommand{\Hk}{H_{\bm k}}
\newcommand{\Eellx}{\frac{\partial \Eell}{\partial k_x}}
\newcommand{\Eelly}{\frac{\partial \Eell}{\partial k_y}}
\newcommand{\Eellpx}{\frac{\partial \Eellp}{\partial k_x}}
\newcommand{\Uellx}{\frac{\partial \Uell}{\partial k_x}}
\newcommand{\Uelly}{\frac{\partial \Uell}{\partial k_y}}
\newcommand{\Uellpx}{\frac{\partial \Uellp}{\partial k_x}}

\newcommand{\Eellm}{\frac{\partial \Eell}{\partial k_\mu}}
\newcommand{\Eelln}{\frac{\partial \Eell}{\partial k_\nu}}
\newcommand{\Eellt}{\frac{\partial \Eell}{\partial k_\tau}}

\newcommand{\Eellpn}{\frac{\partial \Eellp}{\partial k_\nu}}

\newcommand{\Uellm}{\frac{\partial \Uell}{\partial k_\mu}}
\newcommand{\Uelln}{\frac{\partial \Uell}{\partial k_\nu}}

\newcommand{\Uellpm}{\frac{\partial \Uellp}{\partial k_\mu}}

\newcommand{\Uellpy}{\frac{\partial \Uellp}{\partial k_y}}

\newcommand{\Uellppx}{\frac{\partial \Uellpp}{\partial k_x}}

\newcommand\MSum{k_{\rm B} T\sum_{{\bm k},n}}

\title{Orbital Magnetism of Bloch Electrons I. 
General Formula 
}

\author{
Masao \surname{Ogata}\thanks{E-mail: ogata@phys.s.u-tokyo.ac.jp}
and Hidetoshi \surname{Fukuyama}
}

\inst{\ 
Department of Physics, University of Tokyo, Hongo, Bunkyo-ku, Tokyo 113-0033, Japan \\
$^2$Research Institute for Science and Technology, Tokyo University of Science, 
Shinjuku, Tokyo 162-8601, Japan
} 

\abst{
We derive an exact formula of orbital susceptibility expressed in terms of 
Bloch wave functions, starting from the exact one-line formula by Fukuyama 
in terms of Green's functions.
The obtained formula contains four contributions: 
(1) Landau-Peierls susceptibility, (2) interband contribution, 
(3) Fermi surface contribution, and (4) contribution from occupied states. 
Except for the Landau-Peierls susceptibility, the other three contributions involve 
the crystal-momentum derivatives of Bloch wave functions. 
Physical meaning of each term is clarified. 
The present formula is simplified compared with those obtained previously 
by Hebborn {\it et al}. 
Based on the formula, it is seen first of all that diamagnetism from core 
electrons and Van Vleck susceptibility are the only contributions in the atomic limit. 
The band effects are then studied in terms of linear combination of atomic orbital 
treating overlap integrals between atomic orbitals as a perturbation 
and the itinerant feature of Bloch electrons in solids are clarified 
systematically for the first time.
}

\begin{document}
\maketitle

\section{Introduction}

The effect of a magnetic field on electrons in crystals is fascinating and one of the 
basic problems of solid state physics. 
Although the fundamental principles are simple, our understanding of this 
problem has been far from complete
due to the complicated matrix elements between different Bloch bands, 
called as the interband effects of a magnetic field.\cite{Kubo} 
One of the typical problems of this interband effect is the orbital magnetism. 

After the pioneering work of orbital magnetism 
by Landau for free electrons,\cite{Landau}
the effect of periodic potential 
was considered by Peierls\cite{Peierls}
who obtained the 
Landau-Peierls formula for the orbital susceptibility,
\begin{equation}
\chi_{\rm LP} = \frac{e^2}{6\hbar^2 c^2} \sum_{\ell,{\bm k}} \left\{
\frac{\partial^2 \varepsilon_\ell}{\partial k_x^2} 
\frac{\partial^2 \varepsilon_\ell}{\partial k_y^2} - \left(
\frac{\partial^2 \varepsilon_\ell}{\partial k_x \partial k_y}\right)^2   
\right\}
\frac{\partial f(\varepsilon_\ell)}{\partial \varepsilon},
\label{LandauPeierls}
\end{equation}
where $f(\varepsilon)$ is the Fermi distribution function and 
$\varepsilon_\ell ({\bm k})$ is the Bloch band energy. 
This Landau-Peierls formula is obtained by considering the effect 
of a magnetic field by a phase factor
\begin{equation}
{\rm exp}\left( \frac{ie}{\hbar c}\int_{{\bm r}_i}^{{\bm r}_j} {\bm A}\cdot d{\bm \ell}\right),
\label{PeierlsPh}
\end{equation}
(the so-called Peierls phase) in the hopping integral of the single-band 
tight-binding model.\cite{Peierls} 
Here ${\bm A}$ is a vector potential and $e<0$ is the electron charge. 
Attaching the Peierls phase to the hopping integral 
corresponds to the modification of the energy dispersion, 
$\Eell({\bm k}) \rightarrow \Eell({\bm k}-e{\bm A}/c\hbar)$, 
in the presence of a magnetic field. 

Apparently $\chi_{\rm LP}$ does not include the deformation of 
the wave function resulting from the interband matrix elements of the 
magnetic field. 
Therefore it is believed that $\chi_{\rm LP}$ includes only the intraband 
effects.  (It is not so simple as we show later in the present paper.)
Furthermore,  $\chi_{\rm LP}$ vanishes for insulators since it is 
proportional to $\partial f(\varepsilon_\ell)/\partial \varepsilon$.
On the other hand, in bismuth and its alloys, it has been known experimentally 
that the diamagnetism takes its 
maximum when the chemical potential is located in the 
band gap,\cite{BiExp1,BiExp2,BiExp3,BiExp4}
i.e., in the insulating state. 
Apparently the Landau-Peierls formula fails to explain the 
large diamagnetism in bismuth and its alloys, 
which had been a mystery for a long time. 

Stimulated by this experimental fact, there were various efforts to clarify the 
interband effects of magnetic field on orbital susceptibility.%
\cite{Wilson,Adams,Kjeldaas,Roth,Blount,Wannier,Ichimaru,YamajiKubo,HFKubo,HS1,HS2,HS3}
For example, several theoretical studies showed that the difference between 
the total susceptibility and $\chi_{\rm LP}$ is in the same order of the difference 
between $\chi_{\rm LP}$ and Landau susceptibility,\cite{Adams,HFKubo,YamajiKubo} 
even in the nearly-free electron cases. 
The large diamagnetism of bismuth was finally understood by Fukuyama and Kubo\cite{HFKubo2} 
who calculated the magnetic susceptibility based on the Wigner representation. 
It was clarified that the interband effect of a magnetic field and the strong spin-orbit 
interaction are essential. 

After these theoretical efforts, one of the present authors\cite{Fukuyama} 
(hereafter referred as I) discovered an exact but very simple formula of orbital susceptibility as
\begin{equation}
\chi = \frac{e^2}{\hbar^2c^2} k_{\rm B} T \sum_{{\bm k},n} {\rm Tr} \ 
\gamma_x {\cal G} \gamma_y {\cal G} \gamma_x {\cal G} \gamma_y {\cal G},
\label{FukuyamaF}
\end{equation}
where $\cal G$ is the thermal Green's function ${\cal G}({\bm k}, \varepsilon_n)$ 
in a matrix form whose $(ij)$ component is the matrix element between the $i$- and 
$j$-th band. 
$\varepsilon_n$ is Matsubara frequency 
and $\gamma_\mu$ represents the current operator in the 
$\mu$-direction divided by $e/\hbar$. 
The spin multiplicity of 2 has been taken into account and Tr 
is to take trace over the band indices. 
Originally this formula is derived based on the 
Luttinger-Kohn representation.\cite{LuttingerKohn} 
However, as discussed in I, this formula is valid 
in the usual Bloch representation because the two representations are 
related by a unitary transformation and the trace is invariant 
under the unitary transformation. 
This exact one-line formula (\ref{FukuyamaF}) has been applied 
to practical models such as
Weyl equation realized in graphene and an organic conductor 
$\alpha$-(BEDT-TTF)$_2$I$_3$,\cite{FukuGra,Kobayashi,FukuRev}
and Dirac equation in bismuth,\cite{FOF1,FOF2,FOF3,BiRev}
expressed in the Luttinger-Kohn-type Hamiltonians.

For the Bloch representation, the exact formula 
written in terms of Bloch wave functions had been 
derived by Hebborn {\it et al.}\cite{HS1,HS2,HS3} 
(especially in Ref.\cite{HS3} which will be called as HLSS in the following) 
before the exact one-line formula (\ref{FukuyamaF}) was derived. 
It was proved in I with the help of the formulation by Ichimaru\cite{Ichimaru} 
that (\ref{FukuyamaF}) is equivalent to the results by HLSS. 
However HLSS's result is very complicated for the practical use.
Because of this difficulty, orbital susceptibility for Bloch 
electrons in general has not been explored in detail. 
In particular, quantitative estimation of various 
contributions and their physical meaning have not been clarified.

In the single-band tight-binding model, there is a fundamental problem. 
When one restricts the band indices of the Green's functions in (\ref{FukuyamaF}) 
to a single band, one obtains a susceptibility defined as $\chi_1$  
which is different from $\chi_{\rm LP}$.\cite{Fukuyama,KoshinoAndo,Gomez,Piechon}
On the other hand, in the two-dimensional honeycomb lattice 
(or graphene)\cite{Safran,Safran2,Saito,KoshinoAndo,Gomez,Piechon} 
which is a typical two-band tight-binding model, 
it was shown that the orbital susceptibility based on the Peierls phase 
(eq.~(\ref{PeierlsPh})) is not equal to either 
$\chi_1$ or $\chi_{\rm LP}$.\cite{KoshinoAndo,Gomez,Piechon}
From these results, it was claimed that the formula (\ref{FukuyamaF}) cannot be applied 
to the tight-binding models on one hand,\cite{Piechon}  and 
that there are some \lq\lq correction terms'' to the exact formula (\ref{FukuyamaF}) 
on the other hand,\cite{KoshinoAndo,Gomez} both of which are of course unjustified. 
As shown in the present paper, these confusions come from the misusage 
of the exact formula (\ref{FukuyamaF}).

In this paper, starting from the exact one-line formula (\ref{FukuyamaF}) 
and rewriting it in terms of Bloch wave functions, 
we derive a new and exact formula of the orbital susceptibility  
in a different way from those of HLSS.
As shown later explicitly, the new formula is equivalent to the previous 
results.\cite{HS1,HS2,HS3} 
However, it is simpler than the previous results and contains only four contributions: 
(1) Landau-Peierls susceptibility, $\chi_{\rm LP}$, (2) interband contribution, $\chi_{\rm inter}$, 
(3) Fermi surface contribution, $\chi_{\rm FS}$, and (4) contribution from occupied states, $\chi_{\rm occ}$. 
Except for $\chi_{\rm LP}$, the other three contributions involve 
the crystal-momentum derivatives of Bloch wave functions. 
The physical meaning of each term is discussed. 
In the atomic limit, $\chi_{\rm inter}$ is equal to the Van Vleck susceptibility 
and $\chi_{\rm occ}$ is equal to the atomic diamagnetism (or contributions from core-level electrons). 
$\chi_{\rm FS}$ is a newly found contribution proportional to $f'(\Eell)$. 
Then, we apply the present formula to the model of linear combination of 
atomic orbitals. 
We will show that the orbital susceptibility can be calculated systematically by studying the 
effects of overlap integrals between atomic orbitals as a perturbation from the atomic limit. 
In this method, itinerant features of Bloch electrons in solids are clarified for the first time. 
In most of researches, the atomic diamagnetism and Van Vleck contributions are 
treated separately from $\chi_{\rm LP}$. 
However, the present exact formula contains all the contributions on the same basis. 
Furthermore we find that $\chi_{\rm occ}$ contains the contributions not only from the
core-level electrons (known as atomic diamagnetism), 
but also from the occupied states in the partially-filled band. 
This contribution has not been recognized before. 

As mentioned above, when we restrict the band indices of the 
Green's functions in (\ref{FukuyamaF}) to a single band, we do not obtain $\chi_{\rm LP}$. 
In this paper, we show that the total of several contributions in (\ref{FukuyamaF}) 
gives $\chi_{\rm LP}$. 
In these contributions, the $f$-sum rule which involves the summation over the other bands 
plays important roles.  
This means that the band indices of the Green's functions in (\ref{FukuyamaF}) 
should not be restricted to a single band when one consider a 
single-band tight-binding model, 

While preparing this paper, we notice that Gao {\it et al}\cite{Gao}
studied orbital magnetism in terms of Berry phase using the 
wave-packet approximation. 
Their main interest is in the case with broken time-reversal 
symmetry\cite{Niu,Thon1,Niu2,Thon2} which is not the subject of the 
present paper. 
However we can compare our results with theirs in the case where the 
time-reversal symmetry is not broken. 
We find that their results are almost equivalent with ours except for a 
term which has a different prefactor, 
possibly due to the wave-packet approximation they used.

In section 2 we derive a new formula for orbital susceptibility in Bloch 
representation.
Our main results are summarized in eqs.~(\ref{FinalChi})-(\ref{ChiC})
where four contributions are identified. 
In section 3 we apply the obtained formula to the model of linear combination
of atomic orbitals. 
Section 4 is devoted to discussions and future problems. 

\section{Orbital susceptibility in terms of Bloch wave functions}

\subsection{Bloch wave functions and current operator}

In order to explore the implications of eq.~(\ref{FukuyamaF}) in the 
Bloch representation, some essential ingredients are introduced. 
Thermal Green's function for the $\ell$-th band in (\ref{FukuyamaF})
is simply given by 
\begin{equation}
{\cal G}_\ell = \frac{1}{i\varepsilon_n - \Eell({\bm k})}, 
\end{equation}
where $\varepsilon_n$ is Matsubara frequency and $\Eell({\bm k})$ is the Bloch band energy. 
$\ell$ denotes the band index and the wave vector $\bm k$ is within the first 
Brillouin zone. 
In order to obtain the explicit form of the current operator $\gamma_\mu$ in (\ref{FukuyamaF}), 
it is necessary to have information of Bloch wave functions in a periodic potential $V({\bm r})$. 
From the Bloch's theorem, the eigenfunctions of the Hamiltonian are given by
\begin{equation}
e^{i{\bm k}\cdot {\bm r}} \Uell({\bm r}),
\label{Bloch}
\end{equation}
where $\Uell({\bm r})$ is a periodic function with the same 
period as $V({\bm r})$ and satisfies the equation
\begin{equation}
H_{\bm k} \Uell({\bm r}) = \Eell({\bm k}) \Uell({\bm r}),
\label{UellEq}
\end{equation}
with 
\begin{equation}
H_{\bm k} = \frac{\hbar^2 k^2}{2m} - \frac{i\hbar^2}{m} {\bm k}\cdot {\bm \nabla}
- \frac{\hbar^2}{2m} {\bm \nabla}^2 + V({\bm r}).
\label{HamiltonianK}
\end{equation}
Here ${\bm k}\cdot {\bm \nabla}$ indicates the inner product between 
${\bm k}$ and ${\bm \nabla}$.
For simplicity, we assume a centrosymmetric potential, $V(-{\bm r})=V({\bm r})$. 
In this case we can choose $u^\dagger_{\ell {\bm k}}({\bm r}) = u_{\ell {\bm k}}(-{\bm r})$\cite{HS3}
where  $u^\dagger_{\ell {\bm k}}({\bm r})$ is the complex conjugate of 
$u_{\ell {\bm k}}({\bm r})$. 

For $\gamma_\mu$, we calculate the current operator 
\begin{equation}
{\bm j}({\bm r}) = -\frac{ie\hbar }{2m} \sum_\alpha \left[ \psi_\alpha^\dagger ({\bm r}) 
{\bm \nabla} \psi_\alpha ({\bm r}) 
- {\bm \nabla}\psi_\alpha^\dagger ({\bm r}) \ \psi_\alpha ({\bm r}) \right]. 
\label{current_def}
\end{equation}
Substituting the expansion
\begin{equation}
\psi_\alpha ({\bm r}) = \sum_{\ell, {\bm k}} {\hat c}_{\ell {\bm k}\alpha} 
e^{i{\bm k}\cdot {\bm r}} u_{\ell{\bm k}} ({\bm r}), 
\label{PsiExpand}
\end{equation}
into eq.~(\ref{current_def}) and making the Fourier transform, we obtain
\begin{equation}
\begin{split}
{\bm j}_{{\bm q}=0} 
&= \frac{e\hbar}{m} \sum_{\ell \ell' {\bm k} \alpha}
\left[ \int \Uell^\dagger \left( {\bm k} - i{\bm \nabla} \right) \Uellp d{\bm r} \right]
{\hat c}_{\ell {\bm k}\alpha}^\dagger  {\hat c}_{\ell' {\bm k}\alpha} \cr
&= \frac{e}{\hbar} \sum_{\ell \ell' {\bm k} \alpha}
\left[ \int \Uell^\dagger \frac{\partial \Hk}{\partial {\bm k}} \Uellp d{\bm r} \right]
{\hat c}_{\ell {\bm k}\alpha}^\dagger  {\hat c}_{\ell' {\bm k}\alpha},
\label{JFourier}
\end{split}
\end{equation}
where the definition of $\Hk$ in (\ref{HamiltonianK}) has been used. 
Since $\gamma_\mu$ in (\ref{FukuyamaF}) is defined as a current operator 
divided by $e/\hbar$, 
the matrix element of $\gamma_\mu$ is given by (see Appendix A)
\begin{equation}
\left[ \gamma_\mu \right]_{\ell\ell'} 
= \int \Uell^\dagger \frac{\partial \Hk}{\partial k_\mu } \Uellp d{\bm r} 
= \frac{\partial \Eell({\bm k})}{\partial k_\mu } \delta_{\ell \ell'} + p_{\ell \ell' \mu}, 
\label{Jmatrixelement}
\end{equation}
with $p_{\ell \ell' \mu }$ being the off-diagonal matrix elements\cite{WilsonText,Fukuyama}
\begin{equation}
p_{\ell \ell' \mu } = (\Eellp ({\bm k}) - \Eell ({\bm k}) ) 
\int \Uell^\dagger \frac{\partial \Uellp}{\partial k_\mu } d{\bm r}.
\label{Joffdiagonal}
\end{equation}
Although the integral in (\ref{Joffdiagonal}) is sometimes called 
(interband) \lq\lq Berry connection'', this kind of terms has been 
familiar for a long time in the literatures.\cite{Blount2,WilsonText,HS2}
[Note that the intraband \lq\lq Berry connection'' vanishes 
in the present Hamiltonian with $V({\bm r}) = V(-{\bm r})$, 
as shown in (\ref{AppBn2}).]

Here we have used the Fourier integral theorem\cite{LuttingerKohn} 
for functions with the lattice periodicity. 
Originally, the range of the real-space integral on the right-hand side of 
eqs.~(\ref{JFourier}) or (\ref{Joffdiagonal}) is within a unit cell,\cite{LuttingerKohn} 
i.e., $\frac{V}{\Omega}\int_\Omega \cdots d{\bm r}$, where 
$V$ and $\Omega$ are the volumes of the whole system and of the unit cell, respectively. 
However, the range of integral can be extended to the whole system size $V$ 
by using the periodicity of $\Uell ({\bm r})$, 
i.e., $\frac{V}{\Omega}\int_\Omega \cdots d{\bm r}=\int_V \cdots d{\bm r}$. 
In the following, the real-space integrals are defined in this way. 

\subsection{New formula for orbital susceptibility}

Using the above matrix elements for $\gamma_\mu$ and thermal Green's functions, 
we calculate the formula (\ref{FukuyamaF}) in the Bloch representation.
Due to the existence of two terms in each $\left[ \gamma_\mu \right]_{\ell\ell'} $, 
there appear sixteen terms.  
Classifying by the band indices of four Green's functions in (\ref{FukuyamaF}), we obtain
\begin{equation}
\chi = \sum_{n=1}^7 \chi_n,
\end{equation}
with 
\begin{equation}
\chi_1 = \frac{e^2}{\hbar^2 c^2} \MSum \sum_{\ell} \left( \Eellx \right)^2 \left( \Eelly \right)^2  {\cal G}_\ell^4,
\end{equation}
\begin{equation}
\chi_2 =  \frac{2e^2}{\hbar^2 c^2} \MSum \sum_{\ell \ne \ell'} \Eellx \Eelly
p_{\ell\ell' x}p_{\ell' \ell y} \ {\cal G}_\ell^3 {\cal G}_{\ell '} + (x \leftrightarrow y),
\end{equation}
\begin{equation}
\chi_3 =  \frac{e^2}{\hbar^2 c^2} \MSum  \sum_{\ell \ne \ell'}  \Eellx  \Eellpx  
p_{\ell\ell' y}p_{\ell' \ell y} \ {\cal G}_\ell^2 {\cal G}_{\ell '}^2 + (x \leftrightarrow y),
\end{equation}
\begin{equation}
\chi_4 =  \frac{2e^2}{\hbar^2 c^2} \MSum \sum_{\ell \ell' \ell'' }^{'}  \Eellx
p_{\ell\ell' y} p_{\ell' \ell'' x} p_{\ell'' \ell y}  \ {\cal G}_\ell^2 
{\cal G}_{\ell '} {\cal G}_{\ell ''} + (x \leftrightarrow y),
\end{equation}
\begin{equation}
\chi_5 =  \frac{e^2}{\hbar^2 c^2} \MSum \sum_{\ell \ne \ell' }   
p_{\ell\ell' x} p_{\ell' \ell y} p_{\ell \ell' x}  p_{\ell' \ell y} \ {\cal G}_\ell^2 {\cal G}_{\ell '}^2,
\end{equation}
\begin{equation}
\chi_6 =  \frac{e^2}{\hbar^2 c^2} \MSum  \sum_{\ell \ell' \ell''}^{'} 
p_{\ell\ell' x} p_{\ell' \ell y} p_{\ell \ell'' x}  p_{\ell'' \ell y} 
\ {\cal G}_\ell^2 {\cal G}_{\ell '}  {\cal G}_{\ell ''}  + (x \leftrightarrow y),
\end{equation}
\begin{equation}
\chi_7 = \frac{e^2}{\hbar^2 c^2} \MSum \sum_{\ell \ell' \ell'' \ell'''}^{'}   
p_{\ell\ell' x} p_{\ell' \ell'' y} p_{\ell'' \ell''' x}  p_{\ell''' \ell y} 
\ {\cal G}_\ell {\cal G}_{\ell '} {\cal G}_{\ell ''} {\cal G}_{\ell '''},
\end{equation}
where $+(x \leftrightarrow y)$ means a term obtained by replacing $(x,y)$ to $(y,x)$. 
Schematic representation of the contributions to 
 $\chi_1$-$\chi_7$ are shown in Fig.~\ref{Fig:01}. 
The summation with prime $\sum^{'}$ means that all the band indices 
($\ell, \ell', \ell''$ or $\ell, \ell', \ell'', \ell'''$) are different with each other.
In the following we write $\Eell$ for $\Eell({\bm k})$ as far as it is not confusing. 

\def\MC#1{\textcolor{blue}{#1}}
\def\MCC#1{\textcolor{red}{#1}}

\begin{figure}
\setlength{\unitlength}{1mm}
\centering\begin{picture}(80,140)
 \thicklines
  \put(7,130){\Large $\chi_1$} \put(0,130){(a)} 
  \put(20,130){\line(1,0){7}} \put(22,125){$\ell$}  \put(30,130){\line(1,0){7}} \put(32,125){$\ell$}  
  \put(40,130){\line(1,0){7}} \put(42,125){$\ell$}  \put(50,130){\line(1,0){7}} \put(52,125){$\ell$}  
  \put(27.25,129){\textcolor{blue}{$\otimes$}}  \put(37.24,129){\textcolor{blue}{$\otimes$}}
  \put(47.25,129){\textcolor{blue}{$\otimes$}}  \put(57.24,129){\textcolor{blue}{$\otimes$}}
  \put(26,125){\textcolor{blue}{$\frac{\partial \Eell}{\partial k_x}$}}
  \put(7,110){\Large $\chi_2$} \put(0,115){(b)} 
  \put(20,110){\line(1,0){7}} \put(22,105){$\ell$}  \put(30,116){\line(1,0){7}} \put(32,111){$\ell'$} 
  \put(40,110){\line(1,0){7}} \put(44,105){$\ell$}  \put(50,110){\line(1,0){7}} \put(52,105){$\ell$}  
  \MC{\multiput(27,110)(1,2){4}{\circle{1}} \multiput(40,110)(-1,2){4}{\circle{1}}
  \put(21.5,115){$p_{\ell \ell' x}$} \put(40.5,115){$p_{\ell' \ell y}$} }
  \put(47.25,109){\textcolor{blue}{$\otimes$}}  \put(57.24,109){\textcolor{blue}{$\otimes$}}
  \put(7,90){\Large $\chi_3$} \put(0,95){(c)} 
  \put(20,90){\line(1,0){7}} \put(22,85){$\ell$}  \put(30,96){\line(1,0){7}} \put(32,91){$\ell'$} 
  \put(40,96){\line(1,0){7}} \put(42,91){$\ell'$}  \put(50,90){\line(1,0){7}} \put(52,85){$\ell$}  
  \MC{\multiput(27,90)(1,2){4}{\circle{1}} \multiput(50,90)(-1,2){4}{\circle{1}} \put(21.5,95){$p_{\ell \ell' y}$} \put(49.5,95){$p_{\ell' \ell y}$} }
  \put(37.24,95){\textcolor{blue}{$\otimes$}} \put(57.24,89){\textcolor{blue}{$\otimes$}}
  \put(7,70){\Large $\chi_4$} \put(0,75){(d)} 
  \put(20,70){\line(1,0){7}} \put(22,65){$\ell$}  \put(30,76){\line(1,0){7}} \put(32,71){$\ell'$} 
  \put(40,80){\line(1,0){7}} \put(42,75){$\ell''$}  \put(50,70){\line(1,0){7}} \put(52,65){$\ell$}  
  \MC{\multiput(27,70)(1,2){4}{\circle{1}} \multiput(50,70)(-0.6,2){6}{\circle{1}}  \multiput(37,76)(1.8,2){3}{\circle{1}}
  \put(21.5,75){$p_{\ell \ell' y}$} \put(49.5,77){$p_{\ell'' \ell y}$} \put(32,80){$p_{\ell' \ell'' x}$} } 
  \put(57.24,69){\textcolor{blue}{$\otimes$}}
  \put(7,50){\Large $\chi_5$} \put(0,55){(e)} 
  \put(20,50){\line(1,0){7}} \put(22,46){$\ell$}  \put(30,56){\line(1,0){7}} \put(32,51){$\ell'$} 
  \put(40,50){\line(1,0){7}} \put(45,46){$\ell$}  \put(50,56){\line(1,0){7}} \put(52,51){$\ell'$}  
  \MC{\multiput(27,50)(1,2){4}{\circle{1}} \multiput(40,50)(-1,2){4}{\circle{1}} \put(21.5,55){$p_{\ell \ell' x}$} \put(39.5,55){$p_{\ell' \ell y}$}
  \multiput(47,50)(1,2){4}{\circle{1}} \multiput(60,50)(-1,2){4}{\circle{1}} \put(44,58){$p_{\ell \ell' x}$} \put(59.5,55){$p_{\ell' \ell y}$} }
  \put(7,25){\Large $\chi_6$} \put(0,33){(f)} 
  \put(20,25){\line(1,0){7}} \put(22,21){$\ell$}  \put(30,31){\line(1,0){7}} \put(32,26){$\ell'$} 
  \put(40,25){\line(1,0){7}} \put(45,21){$\ell$}  \put(50,35){\line(1,0){7}} \put(52,30){$\ell''$}  
  \MC{\multiput(27,25)(1,2){4}{\circle{1}} \multiput(40,25)(-1,2){4}{\circle{1}} \put(21.5,30){$p_{\ell \ell' x}$} \put(39.5,30){$p_{\ell' \ell y}$}
  \multiput(47,25)(0.6,2){6}{\circle{1}} \multiput(60,25)(-0.6,2){6}{\circle{1}} \put(42,35){$p_{\ell \ell'' x}$} \put(59,35){$p_{\ell'' \ell y}$} }
  \put(7,2){\Large $\chi_7$} \put(0,10){(g)} 
  \put(20,2){\line(1,0){7}} \put(22,-2){$\ell$}  \put(30,8){\line(1,0){7}} \put(32,3){$\ell'$} 
  \put(50,5){\line(1,0){7}} \put(52,0){$\ell'''$}  \put(40,12){\line(1,0){7}} \put(42,7){$\ell''$}  
  \MC{\multiput(27,2)(1,2){4}{\circle{1}} \multiput(37,8)(1.8,2){3}{\circle{1}} \put(21.5,7){$p_{\ell \ell' x}$} \put(32,12){$p_{\ell' \ell'' y}$} 
  \multiput(50,5)(-1,2){4}{\circle{1}} \multiput(60,2)(-1.5,1.5){3}{\circle{1}} \put(50,10){$p_{\ell'' \ell''' x}$} \put(59,5){$p_{\ell''' \ell y}$} }
   \MCC{\put(24,105){\line(1,0){18}} \put(24,120){\line(1,0){18}} \put(24,105){\line(0,1){15}} \put(42,105){\line(0,1){15}}
            \put(24,45){\line(1,0){18}} \put(24,60){\line(1,0){18}} \put(24,45){\line(0,1){15}} \put(42,45){\line(0,1){15}} 
            \put(43,45){\line(1,0){18}} \put(43,60){\line(1,0){18}} \put(43,45){\line(0,1){15}} \put(61,45){\line(0,1){15}}
            \put(24,20){\line(1,0){18}} \put(24,38){\line(1,0){18}} \put(24,20){\line(0,1){18}} \put(42,20){\line(0,1){18}}  
            \put(43,20){\line(1,0){18}} \put(43,38){\line(1,0){18}} \put(43,20){\line(0,1){18}} \put(61,20){\line(0,1){18}} 
            \put(25,101.5){$f$-sum rule} \put(25,41.5){$f$-sum rule} \put(44,41.5){$f$-sum rule}
            \put(25,16.5){$f$-sum rule} \put(44,16.5){$f$-sum rule}
}
\end{picture}
\caption{(Color online) Schematic representation of the contributions to $\chi_1$-$\chi_7$: 
The solid lines with band indices $\ell, \ell'$ etc.\ represent 
the Green's functions. Height of these lines represents the energy level $\Eell$. 
The array of blue circles connecting the two lines represents the off-diagonal matrix elements of 
$\gamma_\mu$, i.e., $p_{\ell\ell', \mu}$. 
The symbol $\otimes$ between the two solid lines represents the diagonal component of 
$\gamma_\mu$, i.e., $\partial \Eell/\partial k_\mu$.
The right-hand of each diagram is connected to its left-hand because of the trace in (\ref{FukuyamaF}). 
The red squares indicate the part of the diagrams which can be expressed by 
the $f$-sum rule in eq.~(\ref{fSumRule}). 
}
\label{Fig:01}
\end{figure}
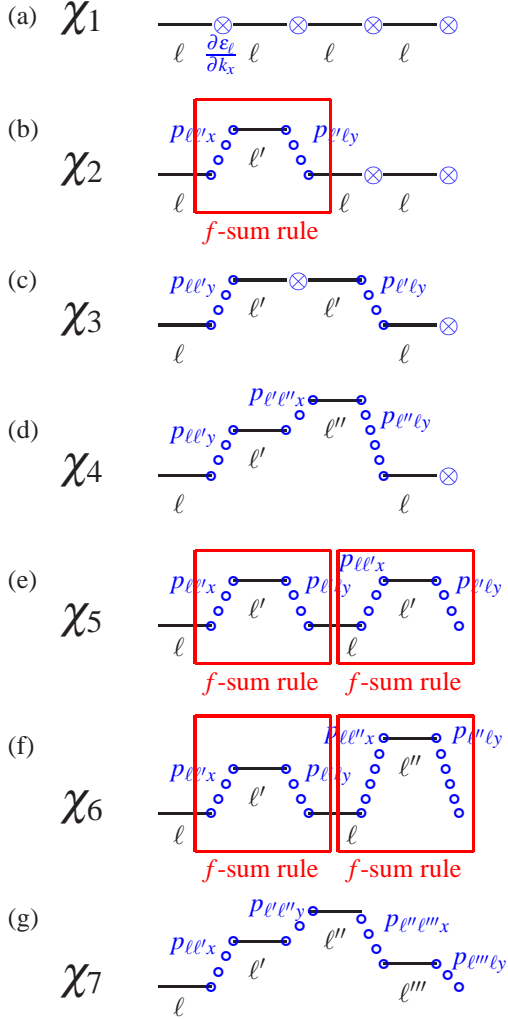

The first contribution $\chi_1$ will be purely intraband since only the intraband matrix 
elements of $\gamma_\mu$'s are involved. 
After taking the summation over Matsubara frequency $n$ and making 
integrations by parts, we obtain\cite{Fukuyama}
\begin{equation}
\begin{split}
\chi_1 &= \frac{e^2}{6\hbar^2 c^2}  \sum_{\ell, {\bm k}} 
\left( \Eellx \right)^2 \left( \Eelly \right)^2 f'''(\Eell) \cr
&= \frac{e^2}{6\hbar^2 c^2} \sum_{\ell, {\bm k}} \biggl[  
\frac{\partial^2 \varepsilon_\ell}{\partial k_x^2} 
\frac{\partial^2 \varepsilon_\ell}{\partial k_y^2} + 2 \left(
\frac{\partial^2 \varepsilon_\ell}{\partial k_x \partial k_y} \right)^2  \cr
&\qquad\ + \frac{3}{2} \left( 
\Eellx \frac{\partial^3 \varepsilon_\ell}{\partial k_x \partial k_y^2} + 
\Eelly \frac{\partial^3 \varepsilon_\ell}{\partial k_x^2 \partial k_y}
\right) \biggr] f'(\varepsilon_\ell), 
\label{Chi1}
\end{split}
\end{equation}
where $f(\Eell)$ is the Fermi distribution function.
This $\chi_1$ is similar to the Landau-Peierls susceptibility $\chi_{\rm LP}$ in (\ref{LandauPeierls}), 
but there are two differences.
The numerical prefactor of the second term of  $\chi_1$ is different from $\chi_{\rm LP}$ and 
the last term of $\chi_1$ does not appear in $\chi_{\rm LP}$.  
We will show shortly that $\chi_{\rm LP}$ is obtained by adding some other 
contributions from  $\chi_2, \chi_5$ and $\chi_6$. 
As discussed in Section 1, this means that one should not pick up only $\chi_1$ in discussing the 
orbital susceptibility in a single-band model. 

Next let us consider $\chi_2$. The summation over $n$ in $\chi_2$ gives 
\begin{equation}
\begin{split}
\chi_2 &= \frac{2e^2}{\hbar^2 c^2} \sum_{\ell \ne \ell', {\bm k}} 
\Eellx \Eelly  p_{\ell\ell' x}p_{\ell' \ell y} \cr
&\times \left\{ \frac{1}{2} \frac{ f''(\Eell)}{\Eell - \Eellp} - \frac{ f'(\Eell)}{(\Eell - \Eellp)^2} 
+ \frac{ f(\Eell) - f(\Eellp) }{(\Eell - \Eellp)^3}  \right\} + (x \leftrightarrow y) \cr
&\equiv \MyChi{2}{1} + \MyChi{2}{2} + \MyChi{2}{3}, 
\label{chi2}
\end{split}
\end{equation}
where the $j$-th term in $\chi_n$ is denoted as $\MyChi{n}{j}$. 
For $\MyChi{2}{1}$, 
the summation over $\ell'$ can be carried out and we obtain
\begin{equation}
\begin{split}
\MyChi{2}{1} = - \frac{e^2}{2\hbar^2 c^2} &\sum_{\ell, {\bm k}} 
f'(\Eell) \biggl\{ 
\left( \frac{\partial^2 \Eell}{\partial k_x \partial k_y} \right)^2
+ \Eellx \frac{\partial^3 \Eell}{\partial k_x \partial k_y^2} \biggr\} \cr
&\qquad \qquad + (x \leftrightarrow y),
\label{Chi2-1}
\end{split}
\end{equation}
where we have used the $f$-sum rule\cite{Wilson,Fukuyama}
\begin{equation}
\sum_{\ell' \ne \ell} \frac{p_{\ell\ell' \mu} p_{\ell' \ell \nu}}{\Eell-\Eellp} 
= \frac{1}{2}\left( \frac{\partial^2 \Eell}{\partial k_\mu \partial k_\nu} -
\frac{\hbar^2}{m} \delta_{\mu\nu} \right),
\label{fSumRule}
\end{equation}
with $\mu=x, \nu=y$ and the integration by parts. 
This $\ell'$-summation is schematically shown in Fig.~\ref{Fig:01}(b), 
in which the red square indicates the part of the diagram representing the $f$-sum rule.  

The $f$-sum rule in eq.~(\ref{fSumRule}) results from the completeness 
property of $\Uellp$ [see eq.~(\ref{fSumRuleInA}) in Appendix A].  
(Various formulas used in the present paper are listed in Appendix A.) 
One may call the left-hand side of (\ref{fSumRule}) as \lq\lq interband" since it 
contains the off-diagonal matrix elements of the current operator $p_{\ell\ell', \mu}$. 
On the other hand, the right-hand side of (\ref{fSumRule}) is expressed by a 
single-band property, $\Eell$, 
and as a result, $\MyChi{2}{1}$ looks like an \lq\lq intraband" contribution.  
This indicates that the naive classification of \lq\lq intraband" and \lq\lq interband" 
does not apply. 

For $\MyChi{2}{2}$, the summation 
over $\ell'$ can also be carried out, and we obtain
\begin{equation}
\begin{split}
\MyChi{2}{2} = - \frac{2e^2}{\hbar^2 c^2} &\sum_{\ell, {\bm k}} 
f'(\Eell) \Eellx  \Eelly
\int \frac{\partial \Uell^\dagger}{\partial k_x} 
\frac{\partial \Uell}{\partial k_y}  d{\bm r} 
+ (x \leftrightarrow y),
\label{Chi2-2}
\end{split}
\end{equation}
where we have used 
\begin{equation}
\sum_{\ell' \ne \ell} \frac{p_{\ell\ell' x} p_{\ell' \ell y}}{(\Eell-\Eellp)^2} 
= \int \frac{\partial \Uell^\dagger}{\partial k_x} \frac{\partial \Uell}{\partial k_y}  d{\bm r}. 
\label{fSumRule2}
\end{equation}
(See (\ref{fSumRule2InA}).)

The last term of $\chi_2$ is given by 
\begin{equation}
\begin{split}
\MyChi{2}{3} &= -\frac{2e^2}{\hbar^2 c^2} \sum_{\ell \ne \ell', {\bm k}} \Eellx  \Eelly 
\frac{ f(\Eell) - f(\Eellp)}{\Eell - \Eellp}  \cr 
&\times \int \Uell^\dagger \Uellpx d{\bm r} \int \Uellp^\dagger \Uelly d{\bm r} 
+ (x \leftrightarrow y). 
\label{Chi2-3}
\end{split}
\end{equation}
Here the $\ell'$-summation can not be carried out due to the presence of the 
denominator, $1/(\Eell - \Eellp)$. 
This denominator is the 
same as what appears in the second-order perturbation of the interband process. 

Features similar to $\chi_2$ are present in $\chi_5$ and $\chi_6$, as seen in 
Fig.\ref{Fig:01}, where the excluded terms of $\ell'=\ell''$ in $\chi_6$ are 
supplemented by $\chi_5$, leading to the independent summations over  
$\ell'$ and $\ell''$.
As a result, using the $f$-sum rule, we obtain
\begin{equation}
\begin{split}
&\MyChi{5}{1} + \MyChi{6}{1} = 
\frac{e^2}{4\hbar^2 c^2} \sum_{\ell, {\bm k}}  f'(\Eell) 
\left( \frac{\partial^2 \Eell}{\partial k_x \partial k_y} \right)^2 + (x \leftrightarrow y). 
\label{Chi56X}
\end{split}
\end{equation}
Detailed calculations are shown in Appendix B. 
[For the definitions of $\MyChi{5}{1}$ and $\MyChi{6}{1}$, see (\ref{AppD1}) 
and  (\ref{AppD2}).]

Now, we can see that sum of $\chi_1$, $\MyChi{2}{1}$ and 
$\MyChi{5}{1} + \MyChi{6}{1}$ becomes
\begin{equation}
\begin{split}
&\frac{e^2}{6\hbar^2 c^2} \sum_{\ell, {\bm k}} f'(\Eell) 
\biggl\{ \frac{\partial^2 \Eell}{\partial k_x^2} \frac{\partial^2 \Eell}{\partial k_y^2} 
-\left( \frac{\partial^2 \Eell}{\partial k_x \partial k_y} \right)^2 \cr
&\qquad \qquad \quad -
\frac{3}{2} \left( 
\Eellx \frac{\partial^3 \varepsilon_\ell}{\partial k_x \partial k_y^2} + 
\Eelly \frac{\partial^3 \varepsilon_\ell}{\partial k_x^2 \partial k_y}
\right) \biggr\}.
\end{split}
\end{equation}
It is seen that the first two terms give $\chi_{\rm LP}$, while the last term 
can be combined with other contributions after the transformation
\begin{equation}
\begin{split}
\frac{\partial^3 \Eell}{\partial k_x \partial k_y^2} 
&= 2\int \frac{\partial \Uell^\dagger}{\partial k_y} \left( 
\frac{\partial H_{\bm k}}{\partial k_x} - \frac{\partial \Eell}{\partial k_x} \right) 
\frac{\partial \Uell}{\partial k_y} d{\bm r} \cr
&+ 4\int \frac{\partial \Uell^\dagger}{\partial k_x} \left( 
\frac{\partial H_{\bm k}}{\partial k_y} - \frac{\partial \Eell}{\partial k_y} \right) 
\frac{\partial \Uell}{\partial k_y} d{\bm r},
\label{E3Formula}
\end{split}
\end{equation}
which is obtained by putting $\mu\nu\tau$ as $xyy$ in eq.~(\ref{AppBe3_2}) in Appendix A. 

Other terms in $\chi_3$-$\chi_7$ are calculated similarly, whose details are 
shown in Appendix B. 
We obtain the total susceptibility $\chi$ as follows, which is exact 
as eq.~(\ref{FukuyamaF}).
\begin{equation}
\chi = \chi_{\rm LP} + \chi_{\rm inter} + \chi_{\rm FS} + \chi_{\rm occ},
\label{FinalChi}
\end{equation}
with 
\begin{equation}
\chi_{\rm LP} = \frac{e^2}{6 \hbar^2 c^2} 
\sum_{\ell, {\bm k}} f'(\Eell) 
\left\{ \frac{\partial^2 \Eell}{\partial k_x^2} \frac{\partial^2 \Eell}{\partial k_y^2} 
-\left( \frac{\partial^2 \Eell}{\partial k_x \partial k_y} \right)^2 \right\},
\label{ChiLP}
\end{equation}
\begin{equation}
\begin{split}
\chi_{\rm inter} &= -\frac{e^2}{\hbar^2 c^2} \sum_{\ell \ne \ell', {\bm k}} \frac{f(\Eell)}{\Eell - \Eellp} 
\biggl| \int \frac{\partial \Uell^\dagger}{\partial k_x} 
\left( \frac{\partial H_{\bm k}}{\partial k_y} + \frac{\partial \Eell}{\partial k_y} \right) \Uellp d{\bm r} \cr 
&\qquad\qquad - \int \frac{\partial \Uell^\dagger}{\partial k_y} 
\left( \frac{\partial H_{\bm k}}{\partial k_x} + \frac{\partial \Eell}{\partial k_x} \right) \Uellp 
d{\bm r} \biggr|^2, 
\label{ChiInter}
\end{split}
\end{equation}
\begin{equation}
\begin{split}
\chi_{\rm FS} &= \frac{e^2}{\hbar^2 c^2} \sum_{\ell, {\bm k}} f'(\Eell) \biggl\{ 
\Eellx\int \frac{\partial \Uell^\dagger}{\partial k_y} 
\left( \frac{\partial H_{\bm k}}{\partial k_x} + \frac{\partial \Eell}{\partial k_x} \right)
\Uelly d{\bm r} \cr
&\qquad
-\Eellx\int \frac{\partial \Uell^\dagger}{\partial k_x} 
\left( \frac{\partial H_{\bm k}}{\partial k_y} + \frac{\partial \Eell}{\partial k_y} \right)
\Uelly d{\bm r}\biggr\} + (x\leftrightarrow y),
\label{ChiFS}
\end{split}
\end{equation}
\begin{equation}
\begin{split}
\chi_{\rm occ} &= -\frac{e^2}{2\hbar^2 c^2}
\sum_{\ell, {\bm k}} f(\Eell) \biggl\{ 
\frac{\partial^2 \Eell}{\partial k_x \partial k_y} 
\int \frac{\partial \Uell^\dagger}{\partial k_x} \frac{\partial \Uell}{\partial k_y} d{\bm r} \cr
&\qquad +\left( \frac{\hbar^2}{m} - \frac{\partial^2 \Eell}{\partial k_x^2} \right)
\int \frac{\partial \Uell^\dagger}{\partial k_y} \frac{\partial \Uell}{\partial k_y} d{\bm r}
\biggr\} + (x\leftrightarrow y).
\label{ChiC}
\end{split}
\end{equation}
Schematic representation of these contributions is shown in Fig.~\ref{Fig:02}. 
It is worth noting that the separation of $\chi_{\rm FS}$ and $\chi_{\rm occ}$ 
is not unique. 
For example, part of $\chi_{\rm occ}$ can be rewritten into a form with 
$f'(\Eell)$ by integration by parts. 
In the above expression, we have chosen a separation which is simple. 

\subsection{Interpretation of each term}

The first contribution, $\chi_{\rm LP}$, is Landau-Peierls formula.
As shown above, $\chi_{\rm LP}$ comes from the 
contributions of $\chi_1$, $\MyChi{2}{1}$ and $\MyChi{5}{1}+\MyChi{6}{1}$.  
As we have seen, the $f$-sum rule plays important roles in $\chi_{\rm LP}$. 

The second contribution, $\chi_{\rm inter}$, is a purely interband contribution. 
Among the four contributions, only this term involves two bands of $\Eell$ and $\Eellp$. 
In the next section, we show that $\chi_{\rm inter}$ is equal to the Van Vleck 
susceptibility when we consider the atomic limit. 

Note that the energy denominator of $\chi_{\rm inter}$ is the same as 
the case of the $f$-sum rule (\ref{fSumRule}). 
However, there is a clear difference. 
Let us consider a single-band case in which the energy differences between the 
bands are quite large. 
In this case, we have $\Delta E = {\rm min} |\Eell ({\bm k}) - \Eellp({\bm k})|$ 
which satisfies $\Delta E>>$ [band width of $\ell$-th band]. 
Then, the absolute value of $\chi_{\rm inter}$ is less than 
\begin{equation}
\begin{split}
&\sum_{\ell \ne \ell', {\bm k} } \frac{f(\Eell)}{\Delta E} 
\biggl| \int \frac{\partial \Uell^\dagger}{\partial k_x} 
\left( \frac{\partial H_{\bm k}}{\partial k_y} + \frac{\partial \Eell}{\partial k_y} \right) 
\Uellp d{\bm r} \cr
&\qquad\qquad\qquad 
- \int \frac{\partial \Uell^\dagger}{\partial k_y} 
\left( \frac{\partial H_{\bm k}}{\partial k_x} + \frac{\partial \Eell}{\partial k_x} 
\right) \Uellp d{\bm r} \biggr|^2.
\end{split}
\end{equation}
Then the $\ell'$ summation can be carried out using the completeness property of $\Uellp$. 
For example, a typical term can be written as
\begin{equation}
\begin{split}
&\sum_{\ell, {\bm k} } \frac{f(\Eell)}{\Delta E} 
\int \frac{\partial \Uell^\dagger}{\partial k_x} 
\left( \frac{\partial H_{\bm k}}{\partial k_y} + \frac{\partial \Eell}{\partial k_y} \right) 
\left( \frac{\partial H_{\bm k}}{\partial k_y} + \frac{\partial \Eell}{\partial k_y} \right) 
\Uellx d{\bm r}, 
\end{split}
\end{equation}
which is arbitrarily small when $\Delta E$ is large.
This is in a sharp contrast with the $f$-sum rule (\ref{fSumRule}) in which the 
right-hand side is finite when $\Delta E$ is large. 

We call the third contribution in (\ref{FinalChi}) as $\chi_{\rm FS}$ 
(FS stands for \lq\lq Fermi surface"), since it is proportional to $f'(\Eell)$. 
This is a newly found contribution, but its physical meaning is not 
clear at present. However, the factor
\begin{equation}
\frac{\partial H_{\bm k}}{\partial k_\mu} + \frac{\partial \Eell}{\partial k_\mu},
\label{CommonFac}
\end{equation}
in the integral in $\chi_{\rm FS}$ is common with $\chi_{\rm inter}$, 
indicating some close relationship between $\chi_{\rm FS}$ and 
$\chi_{\rm inter}$. 

\begin{figure}
\setlength{\unitlength}{1mm}
\centering
\begin{picture}(80,25)
 \thicklines
  \put(7,30){\ }
\end{picture}
\setlength { \fboxrule } { 1pt }
  \colorbox{white}{\begin{picture}(27,10)(0,0)   \end{picture}}
  \colorbox{yellow}{\begin{picture}(27,3)(0,0)   \end{picture}} 
  \colorbox{white}{\begin{picture}(27,13)(0,0)   \end{picture}}
  \colorbox{yellow}{\begin{picture}(27,5)(0,0)   \end{picture}}
\begin{picture}(80,1)
 \thicklines
  \put(71,23){\large $\chi_{\rm LP}, \chi_{\rm FS}$}   \put(39,35){\large $\chi_{\rm inter}$} 
   \put(69.5,11.5){\textcolor{red}{$\left.
\begin{tabular}{@{}l@{}}
\ \\ \ \\ \ \\ \ \\ \   
\end{tabular} \right\}$}} 
  \put(72,13){\large $\chi_{\rm occ}$}
  \put(5,35){\large $\chi^{({\rm Van\ Vleck})}$}  
  \put(4.5,12){\textcolor{red}{$\left\{ 
\begin{tabular}{@{}l@{}}
\ \\ \ \\ \ \\ \   
\end{tabular} \right.$
}}  
  \put(6,12){\large $\chi^{({\rm atomic\ dia.})}$} 
  \put(40.5,10){\textcolor{blue}{\line(1,0){28.8}}} \put(40.5,3){\textcolor{blue}{\line(1,0){28.8}}}
  \put(40.5,3){\textcolor{blue}{\line(0,1){7}}}   \put(69.3,3){\textcolor{blue}{\line(0,1){7}}}  
  \put(40.5,41){\textcolor{blue}{\line(0,1){7}}}   \put(69.3,41){\textcolor{blue}{\line(0,1){7}}}
  \put(40.5,41){\textcolor{blue}{\line(1,0){28.8}}}   \put(40.5,48){\textcolor{blue}{\line(1,0){28.8}}}
  \put(40.5,23){\textcolor{green}{\line(1,0){28.8}}}   \put(40.5,23.9){\textcolor{green}{\line(1,0){28.8}}} 
  \put(40.5,23.3){\textcolor{green}{\line(1,0){28.8}}}   \put(40.5,23.6){\textcolor{green}{\line(1,0){28.8}}}
  \put(40.5,24.2){\textcolor{green}{\line(1,0){28.8}}}
%
  \put(40.5,30){\textcolor{blue}{\line(1,0){28.8}}} \put(40.5,18.4){\textcolor{blue}{\line(1,0){28.8}}}
  \put(40.5,18.4){\textcolor{blue}{\line(0,1){11.6}}}   \put(69.3,18.4){\textcolor{blue}{\line(0,1){11.6}}}  
%
  \put(7,6){\textcolor{blue}{\line(1,0){25}}}  \put(7,20){\textcolor{blue}{\line(1,0){25}}} 
  \put(7,44){\textcolor{blue}{\line(1,0){25}}}  
  \multiput(-1.4,23.5)(4,0){11} {\line(1,0){2}} \put(-2,21){$\mu$}
  \textcolor{red}{\qbezier(5,21)(1.5,32)(6,43)   
                      \qbezier(5,7)(-1,25)(6,43)   
                      \qbezier(39.5,21)(35.5,32)(39.5,43)   
                      \qbezier(39.5,7)(33.5,25)(39.5,43)   
}   \put(52,26){\large $\Eell$}   \put(52,43.5){\large $\Eellp$}   \put(10,-1){atomic limit} \put(46,-1){Bloch bands}
\end{picture}
\caption{(Color online) Schematic representation of energy levels and contributions in the orbital susceptibility 
in the atomic limit (left-hand side) and in the Bloch bands (right-hand side). 
The dashed line represents the position of the chemical potential $\mu$ 
and the colored parts of the squares indicate the occupied states. 
Note that $\chi_{\rm occ}$ has a contribution not only from the core electrons but also from the 
occupied states in the partially-filled band. }
\label{Fig:02}
\end{figure}
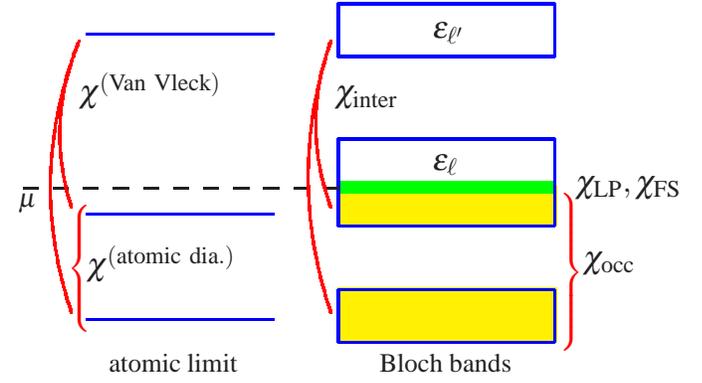

The fourth contribution, $\chi_{\rm occ}$, 
has contributions from the occupied states  (\lq\lq occ" stands for occupied states).  
As shown in the next section, $\chi_{\rm occ}$ is equal to the atomic diamagnetism 
in the atomic limit. 
Furthermore, we show that $\chi_{\rm occ}$ contains the contributions not only from the
core-level electrons, but also from the occupied states in the partially-filled band. 
(See the right-hand side of Fig.~\ref{Fig:02}.)
This contribution has not been recognized before.  

\subsection{Comparison with the result by HLSS}

For the orbital susceptibility in (\ref{FinalChi}), we have only four contributions which are  
simpler than those obtained previously by HLSS, i.e., eqs.~(4.3)-(4.6) in Ref.\cite{HS3}. 
In Appendix C, we prove the equivalence between the present result and HLSS. 
Here we summarize the differences between the two. 

(1) The result by HLSS is not symmetric with respect to the 
exchange of $x$ and $y$. 
This is because they used Landau gauge, ${\bm A}=(-Hy, 0, 0)$. 
On the other hand, the present formula is symmetric with respect to $x$ and $y$ 
because we have used the gauge-invariant formalism (\ref{FukuyamaF}). 
In order to prove the equivalence between our result and that by HLSS, 
we have to symmetrize the HLSS's result. (See details in Appendix C.)

(2) Among the four contributions in the present formula, 
$\chi_{\rm LP}$ is determined solely from the energy dispersion $\Eell({\bm k})$. 
The other three contributions involve the ${\bm k}$-derivatives of wave functions. 
In contrast, HLSS's result contains a term 
\begin{equation}
\frac{e^2}{6\hbar^2 c^2} 
\sum_{\ell, {\bm k}} f'(\Eell) 
\frac{3}{2} \left( \Eellx \frac{\partial^3 \Eell}{\partial k_x \partial k_y^2} 
+ \Eelly \frac{\partial^3 \Eell}{\partial k_x^2 \partial k_y} \right).
\label{HSyobun}
\end{equation}
(See eq.~(\ref{HSChi}).) 
As shown above in eq.~(\ref{E3Formula}), this term can be rewritten 
in terms of $\Uell$ and has been included in $\chi_{\rm FS}$ in our formalism. 
It is important to use (\ref{E3Formula}) in order to simplify the final expression. 
Note that, in contrast to (\ref{HSyobun}), $\chi_{\rm LP}$ can not be 
rewritten in terms of $\Uell$'s. 

(3) The result by HLSS contains several terms which have a 
common denominator of $1/(\Eell-\Eellp)$. 
In our formula, these contributions are summed up into a single 
term as $\chi_{\rm inter}$. 
The method how we can sum up the several terms in HLSS 
into a single term is explained in Appendix C. 

(4) As explained above, each contribution in the present formula has a rather 
clear meaning compared with the previous ones.
For example, $\chi_{\rm inter}$ and $\chi_{\rm occ}$ are contributions naturally 
connected to the Van Vleck susceptibility and atomic diamagnetism, respectively. 
Note that, in the HLSS's formula, 
the contribution of the atomic diamagnetism is expressed as the first term of 
$\chi_4^{({\rm HLSS})}$ in eq.~(\ref{HSChi}), i.e., 
\begin{equation}
-  \frac{2e^2}{\hbar^2 c^2} \sum_{\ell, {\bm k}} \frac{\hbar^2}{m} 
f(\Eell) \int \frac{\partial \Uell^\dagger}{\partial k_y} \Uelly d{\bm r}. 
\label{HScoreterm}
\end{equation}
However, the numerical prefactor is different from that of the present result. 
(See the term proportional to $\hbar^2/m$ in $\chi_{\rm occ}$). 
As shown in the next section, this term reproduces the orbital susceptibility 
from the core-electrons in the atomic limit.  
$\chi_{\rm occ}$ in the present formula gives a correct prefactor, 
while eq.~(\ref{HScoreterm}) by HLSS does not. 
As shown in Appendix C, we find that the correct term  
is obtained in the HLSS's formula when we rewrite the interband contributions into a single 
term as $\chi_{\rm inter}$. 
[For details, see eq.~(\ref{AppE4new}).]

\subsection{Comparison with the results by Gao {\it et al}.}

Let us discuss here the recent work by Gao {\it et al}\cite{Gao}
who studied orbital magnetism in terms of Berry phase. 
They are interested in the case with broken time-reversal 
symmetry in which spontaneous orbital magnetization appears.\cite{Niu,Thon1,Niu2,Thon2}
In this case, there are several terms involving the Berry curvature 
denoted as ${\bm \Omega}$.
In the present notation, its $z$-component is given by 
\begin{equation}
\Omega_z = i\int \left( 
\frac{\partial \Uell^\dagger}{\partial k_x} \frac{\partial \Uell}{\partial k_y}-
\frac{\partial \Uell^\dagger}{\partial k_y} \frac{\partial \Uell}{\partial k_x} 
\right) d{\bm r}. 
\end{equation}
However, in our case with a centrosymmetric potential, 
$\Omega_z$ vanishes since there is a relation 
$\Uell^\dagger(-{\bm r}) = \Uell({\bm r})$. 
As a result, we do not have the contributions coming from the Berry curvature. 
However, we can compare our results with theirs in the case where 
the time-reversal symmetry is not broken. 

Details of calculations are shown in Appendix D. 
By using the completeness property of $\Uell$, we can show that their results 
are almost equivalent with our results except for the coefficient of $\chi_{\rm FS}$.  
We think that this difference is due to the wave-packet approximation\cite{Niu2} 
used in their formalism.
Nevertheless, their formula for the orbital susceptibility 
based on the wave-packet approximation is fairly accurate. 

\section{Band effect from atomic limit}

The obtained formula in eqs.~(\ref{ChiLP})-(\ref{ChiC}) is exact. 
However, in order to calculate each contribution explicitly, it is necessary to specify 
the functional form of $\Uell({\bm r})$. 
For example, $\Uell({\bm r})$ can be obtained in general from the first-principle band calculation. 
In this paper, however, we will study each contribution from the atomic limit 
to see possible band effects based on the linear combination of atomic orbitals (LCAO). 
In the atomic limit, it is found that   
diamagnetic susceptibility from core electrons and Van Vleck susceptibility are 
the only contributions to $\chi$. 
Then, $\chi$ is estimated 
by treating the overlap integrals between atomic orbitals as a perturbation. 
We will show that there appear several contributions to $\chi$
in addition to $\chi_{\rm LP}$.  
In this perturbative method, the itinerant feature of Bloch electrons
in solids are clarified systematically. 
Schematic picture is shown in Fig.~\ref{Fig:02}.

\subsection{Atomic limit}

In order to study the atomic limit in the present formula, it is appropriate to
use LCAO. 
Let us consider a situation in which the periodic potential $V({\bm r})$ is 
written as 
\begin{equation}
V({\bm r}) = \sum_{{\bm R}_i} V_0({\bm r}-{\bm R}_i),
\label{PotSum}
\end{equation}
where ${\bm R}_i$ represent lattice sites and $V_0({\bm r})$ is a potential 
of a single atom. 
We use atomic orbitals $\phi_n({\bm r})$ which satisfy   
\begin{equation}
\left( -\frac{\hbar^2}{2m} \nabla^2 + V_0({\bm r}) \right) \phi_n ({\bm r}) = 
E_n \phi_n ({\bm r}). 
\label{phiEq}
\end{equation}
Using these atomic orbitals, we consider the LCAO wave function
\begin{equation}
\varphi_{n{\bm k}} ({\bm r}) = \frac{1}{\sqrt{N}}
\sum_{{\bm R}_i} e^{-i{\bm k}({\bm r}-{\bm R}_i)} \phi_n ({\bm r}-{\bm R}_i),
\label{LCAO}
\end{equation}
which is used as a basis set for $\Uell({\bm r})$. 
Here $N$ is the total number of unit cells.
It is easily shown that $\varphi_{n{\bm k}}({\bm r})$ are periodic functions with 
the same period with $V({\bm r})$. 

In the atomic limit, 
$V_0({\bm r}-{\bm R}_i)$ and $\phi_n ({\bm r}-{\bm R}_i)$ 
are confined in a unit cell and there is no overlap between nearest-neighbor 
$V_0({\bm r}-{\bm R}_i)$ or nearest-neighbor $\phi_n ({\bm r}-{\bm R}_i)$. 
In this case, it is easily shown that the LCAO wave function, 
$\varphi_{\ell{\bm k}} ({\bm r})$, in (\ref{LCAO}) satisfies the equation (\ref{UellEq}) 
with energy $\Eell=E_\ell$ that is independent on $\bm k$. 
Therefore, $\Uell$ is just given by
\begin{equation}
\Uell ({\bm r}) = \varphi_{\ell{\bm k}} ({\bm r}) = \frac{1}{\sqrt{N}}
\sum_{{\bm R}_i} e^{-i{\bm k}({\bm r}-{\bm R}_i)} \phi_\ell ({\bm r}-{\bm R}_i).
\label{AtomicLim}
\end{equation}

By substituting eq.~(\ref{AtomicLim}) and $\Eell = E_\ell$ 
into eqs.~(\ref{ChiLP})-(\ref{ChiC}), 
we obtain $\chi_{\rm LP}, \chi_{\rm inter}, \chi_{\rm FS}$ and $\chi_{\rm occ}$. 
Since  $E_\ell$ is $\bm k$-independent, $\chi_{\rm LP}=\chi_{\rm FS} = 0$. 
For $\chi_{\rm inter}$, we obtain
\begin{equation}
\begin{split}
&\chi_{\rm inter} = -\frac{e^2}{\hbar^2 c^2} \sum_{\ell \ne \ell', {\bm k}} 
\frac{f(E_\ell)}{E_\ell - E_{\ell'}} \biggl| \frac{1}{N} \frac{\hbar^2}{m}\sum_{{\bm R}_i, {\bm R}_j} \cr
& \int (x-R_{jx}) e^{i{\bm k}({\bm r}-{\bm R}_j)} \phi_\ell^* ({\bm r}-{\bm R}_j)
e^{-i{\bm k}({\bm r}-{\bm R}_i)}  \nabla_y  \phi_{\ell'} ({\bm r}-{\bm R}_i) d{\bm r} \cr
& -\int (y-R_{jy}) e^{i{\bm k}({\bm r}-{\bm R}_j)} \phi_\ell^* ({\bm r}-{\bm R}_j)
e^{-i{\bm k}({\bm r}-{\bm R}_i)} \nabla_x\phi_{\ell'} ({\bm r}-{\bm R}_i) d{\bm r} \biggr|^2.
\end{split}
\end{equation}
Since there is no overlap between atomic orbitals, only the terms with 
${{\bm R}_i={\bm R}_j}$ survives. Thus $\chi_{\rm inter}$ is simplified as
\begin{equation}
\begin{split}
\chi_{\rm inter} &= -\frac{e^2}{\hbar^2 c^2} \sum_{\ell \ne \ell', {\bm k}} 
\frac{f(E_\ell)}{E_\ell - E_{\ell'}} \cr 
&\times \biggl| 
\frac{\hbar^2}{m} \int \phi_\ell^* ({\bm r})  (x\nabla_y - y\nabla_x) \phi_{\ell'} ({\bm r}) 
d{\bm r} \biggr|^2, 
\label{vanVleckProper}
\end{split}
\end{equation}
where the ${\bm r}$-integral is shifted to the center of the atomic orbital, ${\bm R}_i$. 
The right-hand side of (\ref{vanVleckProper}) is nothing but the Van Vleck susceptibility
which we denote as $\chi^{\rm ({\rm Van\ Vleck})}$. 

Similarly, by substituting (\ref{AtomicLim}) and $\Eell=E_\ell$ into (\ref{ChiC}), we obtain
\begin{equation}
\begin{split}
\chi_{\rm occ} &= -\frac{e^2}{2\hbar^2 c^2}
\sum_{\ell, {\bm k}} f(E_\ell) \frac{\hbar^2}{m} 
\int \frac{\partial \varphi_{\ell{\bm k}}^\dagger}{\partial k_x} \frac{\partial \varphi_{\ell{\bm k}}}{\partial k_x} d{\bm r}
+ (x \leftrightarrow y) \cr
&= -\frac{e^2}{2\hbar^2 c^2} \sum_{\ell, {\bm k}} f(E_\ell) \frac{\hbar^2}{m} 
\int (x^2+y^2) |\phi_{\ell} ({\bm r})|^2 d{\bm r}. 
\label{coreProper}
\end{split}
\end{equation}
This is just the atomic diamagnetism coming from the core electrons 
which we denote as $\chi^{\rm ({\rm atomic\ dia.})}$. 
Therefore, in the atomic limit, we have 
$\chi=\chi^{\rm ({\rm Van\ Vleck})}+\chi^{\rm ({\rm atomic\ dia.})}$.  
(See the left-hand side of Fig.~\ref{Fig:02}.)

\subsection{Perturbation with respect to the overlap integrals}

Next we consider the case in which there are overlap integrals 
between the nearest-neighbor atomic orbitals. 
In this case, using $\varphi_{n{\bm k}}({\bm r})$ in eq.~(\ref{LCAO}), 
we expand $\Uell$ as 
\begin{equation}
\begin{split}
\Uell ({\bm r}) &= \sum_n c_{\ell, n}({\bm k}) \varphi_{n{\bm k}} ({\bm r}) \cr
&=\frac{1}{\sqrt{N}} \sum_n \sum_{{\bm R}_i} c_{\ell, n}({\bm k}) 
e^{-i{\bm k}({\bm r}-{\bm R}_i)} \phi_n ({\bm r}-{\bm R}_i).
\label{UellTBA}
\end{split}
\end{equation}
(Note that $c_{\ell, n}({\bm k})=\delta_{\ell, n} $ in the atomic limit.)
The coefficients $c_{\ell, n}({\bm k})$ should be determined in order for $\Uell$ to 
satisfy the equation (\ref{UellEq}).
This can be achieved by solving the eigenvalue problem
\begin{equation}
\sum_m h_{nm}({\bm k}) c_{\ell, m}({\bm k}) = \Eell({\bm k}) \sum_m s_{nm}({\bm k}) c_{\ell, m}({\bm k}),
\label{EigenEquation}
\end{equation}
where the Hamiltonian matrix elements are 
\begin{equation}
h_{nm}({\bm k}) = \int \varphi_{n{\bm k}}^*({\bm r}) H_{\bm k} \varphi_{m{\bm k}}({\bm r}) d{\bm r},
\end{equation}
and $s_{nm}({\bm k})$ represents the integral
\begin{equation}
s_{nm}({\bm k}) = \int \varphi_{n{\bm k}}^*({\bm r}) \varphi_{m{\bm k}}({\bm r}) d{\bm r}.
\end{equation}
$h_{nm}({\bm k})$ and $s_{nm}({\bm k})$ can be calculated perturbatively with respect 
to the overlap integral
\begin{equation}
\int \phi_n^* ({\bm r}-{\bm R}_j) {\cal O}  \phi_m ({\bm r}-{\bm R}_i) d{\bm r},
\end{equation}
with $\cal O$ being an operator and ${\bm R}_j \ne {\bm R}_i$. 
For example, the first-order term of $h_{nm}({\bm k})$ contains 
the hopping integral used in the tight-binding model. 

By substituting eq.~(\ref{UellTBA}) and $\Eell({\bm k})$ into 
eqs.~(\ref{ChiLP})-(\ref{ChiC}), 
we can show that each of four contributions, 
$\chi_{\rm LP}, \chi_{\rm inter}, \chi_{\rm FS}$ and $\chi_{\rm occ}$, 
is calculated perturbatively with respect to the overlap integrals.  
In contrast to the atomic limit, there are two new features: 
(1) $\Eell({\bm k})$ has band dispersion due to the hopping integrals, and 
(2) $\Uell({\bm r})$ has an additional $\bm k$-dependence through 
$c_{\ell,n}({\bm k})$ in eq.~(\ref{UellTBA}). 
The latter gives several contributions to the orbital susceptibility 
originating from the $\bm k$-derivatives of $\Uell({\bm r})$.
One may expect that $\chi_{\rm LP}$ is dominant in the first-order perturbation. 
However, we find that the situation is not so simple even in the single-band case. 
Each of $\chi_{\rm LP}, \chi_{\rm inter}, \chi_{\rm FS}$ 
and $\chi_{\rm occ}$ depends on the location of the chemical potential as well as 
on the details of the model. 
In the forthcoming paper, 
we will discuss several explicit models such as single-band and two-band tight-binding 
models. 

\section{Discussions}

Based on the exact formula, 
we have shown rigorously that the orbital susceptibility for Bloch electrons can be 
described in terms of four contributions, 
$\chi=\chi_{\rm LP}+\chi_{\rm inter}+\chi_{\rm FS}+\chi_{\rm occ}$.
Except for the Landau-Peierls susceptibility, $\chi_{\rm LP}$, 
the other three contributions involve 
the crystal-momentum derivatives of $\Uell$'s. 
These contributions represent the effects of the deformation of the wave function 
due to the magnetic field. 
We find that $\chi_{\rm occ}$ contains the contributions from the occupied states 
in the partially-filled band, which has not been recognized before.  
We applied the present formula to the model of LCAO. 
In the atomic limit where there are no overlap integrals, 
$\chi_{\rm inter}$ becomes $\chi^{\rm ({\rm Van\ Vleck})}$ and 
$\chi_{\rm occ}$ becomes $\chi^{\rm ({\rm atomic\ dia.})}$. 
These two are the only contributions to $\chi$ in the atomic limit. 
When the overlap integrals are finite, we have discussed that $\chi$ can be
calculated by treating the overlap integrals as a perturbation. 
In this method, itinerant features of Bloch electrons in solids can be 
clarified systematically for the first time. 

The present formalism can be used as a starting point for various extensions. 
Several future problems are as follows:

(1) It is very interesting to apply the present formula to 
the multi-band tight-binding models. 
A typical example is the honeycomb lattice which is a model for 
graphene.\cite{Safran,Safran2,Saito,KoshinoAndo,Gomez,Piechon} 
In this case, we have A- and B-sublattice in a unit cell, and as a result, 
we have massless Dirac electrons (or more precisely, Weyl electrons) which is 
a typical two-band model. 
The orbital susceptibility has been calculated by several 
groups\cite{KoshinoAndo,Gomez,Piechon} based on the Peierls phase. 
In contrast, in the present formula, all the contributions from Bloch bands 
are included rigorously. 
Application of the present formula to graphene will be discussed 
in the forthcoming paper.

(2) In the present Hamiltonian, the spin-orbit interaction is not included. 
It is also a very interesting problem to study the orbital susceptibility 
in the presence of spin-orbit interaction. 
As discussed recently by Gao {\it et al.},\cite{Gao}
the orbital susceptibility in the Hamiltonian with broken time-reversal 
symmetry\cite{Niu,Thon1,Niu2,Thon2} is another interesting problem. 
This will be also studied in the forthcoming paper. 
As suggested\cite{Gao} there appear several terms which is written in terms of 
Berry curvatures. 

(3) We have confined ourselves in the orbital susceptibility in this paper.  
The transport coefficients are of course interesting quantities.\cite{FukuyamaHall} 
Hall conductivity in the Weyl equation realized in graphene and an organic conductor 
$\alpha$-(BEDT-TTF)$_2$I$_3$\cite{FukuGra,Kobayashi,FukuRev} as well as in bismuth\cite{FOF1,BiRev}
has been discussed. 
The similar method used in this paper can be applied to the Hall conductivity 
in the Bloch representation. 

\bigskip\noindent
{\bf Acknowledgment}

We thank very fruitful discussions with F.\ Pi\'echon, I.\ Proskurin, Y.\ Fuseya,
H.\ Matsuura, T.\ Mizoguchi and N.\ Okuma. 
This paper is dedicated to Professor Ryogo Kubo (1920-1995), who guided 
authors into the never-fashionable but very deep and fundamentally 
important problem of orbital magnetism in solids.
This work was supported by a Grant-in-Aid for Scientific Research on 
\lq\lq Dirac Electrons in Solids'' (No.\ 24244053) and 
\lq\lq Multiferroics in Dirac electron materials'' (No.\ 15H02108).

\onecolumn

\appendix
\section{Various formulas for matrix elements of Bloch wave functions}

In this appendix, we derive various useful formulas which are quite often used 
in the derivations. 
First, we make ${\bm k}$ derivative of the equation for $\Uell$ in eq.~(\ref{UellEq}) 
\begin{equation}
\frac{\partial H_{\bm k}}{\partial {\bm k}} \Uell + H_{\bm k}  \frac{\partial \Uell}{\partial {\bm k}}
= \frac{\partial \Eell}{\partial {\bm k}} \Uell + \Eell \frac{\partial \Uell}{\partial {\bm k}}.
\label{AppBder1}
\end{equation}
In the following, we write $\Eell$ for $\Eell({\bm k})$. 
When we multiply $\Uellp^\dagger$ and make the real-space integral, we obtain
\begin{equation}
\int \Uellp^\dagger \frac{\partial H_{\bm k}}{\partial {\bm k}} \Uell d{\bm r} 
= \frac{\partial \Eell}{\partial {\bm k}} \delta_{\ell \ell'} + 
(\Eell - \Eellp) \int \Uellp^\dagger \frac{\partial \Uell}{\partial {\bm k}} d{\bm r}.
\label{AppB1p}
\end{equation}
Exchange of $\ell \leftrightarrow \ell'$ gives eq.~(\ref{Jmatrixelement}).

Next, the $k_\mu$ and $k_\nu$ derivative of eq.~(\ref{UellEq}) gives
\begin{equation}
\left( \frac{\hbar^2}{m} \delta_{\mu\nu} -\frac{\partial^2 \Eell}{\partial k_\mu \partial k_\nu}\right) \Uell
+ \left( \frac{\partial H_{\bm k}}{\partial k_\mu} - \frac{\partial \Eell}{\partial k_\mu} \right) \Uelln 
+ \left( \frac{\partial H_{\bm k}}{\partial k_\nu} - \frac{\partial \Eell}{\partial k_\nu} \right) \Uellm  
+ \left( H_{\bm k} - \Eell \right) \frac{\partial^2 \Uell}{\partial k_\mu \partial k_\nu}
= 0. 
\label{AppBder2}
\end{equation}
When we multiply $\Uellp^\dagger$ and make the real-space integral, we obtain
\begin{equation}
\left( \frac{\hbar^2}{m} \delta_{\mu\nu} -  \frac{\partial^2 \Eell}{\partial k_\mu \partial k_\nu} 
\right) \delta_{\ell \ell'} 
+ \int \Uellp^\dagger \left( 
\frac{\partial H_{\bm k}}{\partial k_\mu} - \frac{\partial \Eell}{\partial k_\mu} \right) \Uelln d{\bm r} 
+ \int \Uellp^\dagger \left( 
\frac{\partial H_{\bm k}}{\partial k_\nu} - \frac{\partial \Eell}{\partial k_\nu} \right) \Uellm d{\bm r} 
+ \left( \Eellp - \Eell \right) \int \Uellp^\dagger \frac{\partial^2 \Uell}{\partial k_\mu \partial k_\nu}
d{\bm r} = 0.
\label{AppB3p}
\end{equation}
Similarly, when we multiply $\partial^2\Uell^\dagger/\partial k_\tau \partial k_\sigma$ 
and make the real-space integral, then we obtain
\begin{equation}
\begin{split}
\int \frac{\partial^2\Uell^\dagger}{\partial k_\tau \partial k_\sigma}
(\Eell - H_{\bm k}) \frac{\partial^2 \Uell}{\partial k_\mu \partial k_\nu} d{\bm r} 
&= \int \frac{\partial^2\Uell^\dagger}{\partial k_\tau \partial k_\sigma} \left( 
\frac{\partial H_{\bm k}}{\partial k_\mu} - \frac{\partial \Eell}{\partial k_\mu} \right) \Uelln d{\bm r} 
+ \int \frac{\partial^2\Uell^\dagger}{\partial k_\tau \partial k_\sigma} \left( 
\frac{\partial H_{\bm k}}{\partial k_\nu} - \frac{\partial \Eell}{\partial k_\nu} \right) \Uellm d{\bm r} \cr 
&\qquad +\left( \frac{\hbar^2}{m} \delta_{\mu\nu} -  \frac{\partial^2 \Eell}{\partial k_\mu \partial k_\nu} 
\right) \int \frac{\partial^2\Uell^\dagger}{\partial k_\tau \partial k_\sigma} \Uell d{\bm r}.
\label{AppBx}
\end{split}
\end{equation}

We can obtain other series of formulas by making $k$-derivative of the ortho-normal condition, 
$\int \Uellp^\dagger \Uell d{\bm r} = \delta_{\ell \ell'}$.
First we obtain
\begin{equation}
\int \Uellp^\dagger \Uellm d{\bm r} = - \int \frac{\partial \Uellp^\dagger}{\partial k_\mu} \Uell d{\bm r}.
\label{AppBn1}
\end{equation}
Furthermore if the system is centrosymmetric, i.e., if $V(-{\bm r}) = V({\bm r})$ holds, 
we can choose $\Uell^\dagger (-{\bm r}) = \Uell ({\bm r})$.\cite{HS3}
In this case, we can make further convenient formulas. 
Firstly using the change of variable ${\bm r}\rightarrow -{\bm r}$, we obtain
\begin{equation}
\int \Uell^\dagger \Uellm d{\bm r} = \int \frac{\partial \Uell^\dagger}{\partial k_\mu} \Uell d{\bm r}.
\end{equation}
Combining this relation with (\ref{AppBn1}) with $\ell=\ell'$, we obtain
\begin{equation}
\int \Uell^\dagger \Uellm d{\bm r} = \int \frac{\partial \Uell^\dagger}{\partial k_\mu} \Uell d{\bm r}=0.
\label{AppBn2}
\end{equation}
(This integral is the so-called intraband \lq\lq Berry connection''.)
This formula is used quite often. 
Further $k$-derivatives give
\begin{equation}
\int \Uell^\dagger \frac{\partial^2 \Uell}{\partial k_\mu \partial k_\nu}  d{\bm r}
= - \int \frac{\partial \Uell^\dagger}{\partial k_\nu} \Uellm d{\bm r}.
\label{AppBn3}
\end{equation}
Using the relation $\Uell(-{\bm r}) = \Uell^\dagger ({\bm r})$, we can also show that 
$p_{\ell \ell' \mu}$ in (\ref{Joffdiagonal}) is real and
\begin{equation}
\int \frac{\partial^2 \Uell^\dagger}{\partial k_\tau \partial k_\sigma}
\frac{\partial H_{\bm k}}{\partial k_\mu} \frac{\partial \Uell}{\partial k_\nu} d{\bm r}
= \int \frac{\partial \Uell^\dagger}{\partial k_\nu}
\frac{\partial H_{\bm k}}{\partial k_\mu} \frac{\partial^2 \Uell}{\partial k_\tau \partial k_\sigma} d{\bm r}.
\label{AppBn4}
\end{equation}

Furthermore, from the $k_\nu$ derivative of eq.~(\ref{AppB1p}) with $\ell=\ell'$, we obtain 
\begin{equation}
\int \Uell^\dagger  \frac{\partial H_{\bm k}}{\partial k_\mu} \Uelln d{\bm r} 
= \int \frac{\partial \Uell^\dagger}{\partial k_\nu} \frac{\partial H_{\bm k}} {\partial k_\mu} \Uell d{\bm r}
= \frac{1}{2}\left(\frac{\partial^2 \Eell}{\partial k_\mu \partial k_\nu} 
- \frac{\hbar^2}{m} \delta_{\mu\nu} \right).
\label{AppB1}
\end{equation}
Note that $\mu$ and $\nu$ can be exchanged ($\mu\leftrightarrow \nu$) in (\ref{AppB1}).
Finally, using (\ref{AppB1}) and (\ref{AppBder1}), we have 
\begin{equation}
\int \frac{\partial \Uell^\dagger}{\partial k_\mu} (\Eell- H_{\bm k}) 
\frac{\partial \Uell}{\partial k_\nu}  d{\bm r}
= \int \frac{\partial \Uell^\dagger}{\partial k_\mu} \left( 
\frac{\partial H_{\bm k}}{\partial k_\nu} - \Eelln \right) \Uell d{\bm r}
=  \frac{1}{2}\left(\frac{\partial^2 \Eell}{\partial k_\mu \partial k_\nu} 
- \frac{\hbar^2}{m} \delta_{\mu\nu} \right).
\label{AppBX}
\end{equation}

By using $p_{\ell\ell',\mu}$ in (\ref{Joffdiagonal}), we obtain the $f$-sum rule\cite{Wilson,Fukuyama} as follows:
\begin{equation}
\begin{split}
\sum_{\ell' \ne \ell} \frac{p_{\ell\ell' \mu} p_{\ell' \ell \nu}}{\Eell-\Eellp} 
&= -\sum_{\ell' \ne \ell} (\Eell-\Eellp) \int \Uell^\dagger \Uellpm d{\bm r} \int \Uellp^\dagger \Uelln d{\bm r} 
= \sum_{\ell'} \int \frac{\partial \Uell^\dagger}{\partial k_\mu} (\Eell-H_{\bm k}) \Uellp d{\bm r} 
\int \Uellp^\dagger \Uelln d{\bm r} \cr
&=\int \frac{\partial \Uell^\dagger}{\partial k_\mu} (\Eell- H_{\bm k}) 
\frac{\partial \Uell}{\partial k_\nu}  d{\bm r} 
= \frac{1}{2}\left( \frac{\partial^2 \Eell}{\partial k_\mu \partial k_\nu} -
\frac{\hbar^2}{m} \delta_{\mu\nu} \right),
\label{fSumRuleInA}
\end{split}
\end{equation}
where we have added the $\ell' = \ell$ term which is equal to zero due to the factor
$(\Eell-\Eellp)$. 
We have also used the relations (\ref{AppBder1}) and (\ref{AppBn1}), 
and the completeness property, 
$\sum_{\ell'} \Uellp ({\bm r}) \Uellp^\dagger({\bm r}') = \delta({\bm r}-{\bm r}')$. 
Similarly we obtain
\begin{equation}
\begin{split}
\sum_{\ell'' \ne \ell, \ell'} \frac{p_{\ell\ell'' \mu} p_{\ell'' \ell' \nu}}{\Eell-\Eellpp} 
&=\int \frac{\partial \Uell^\dagger}{\partial k_\mu} (\Eellp - H_{\bm k}) 
\frac{\partial \Uellp}{\partial k_\nu}  d{\bm r} 
=\int \frac{\partial \Uell^\dagger}{\partial k_\mu} \left( \frac{\partial H_{\bm k}}{\partial k_\nu} - \Eellpn\right) 
\Uellp  d{\bm r},
\label{fSumRule3InA}
\end{split}
\end{equation}
\begin{equation}
\sum_{\ell' \ne \ell} \frac{p_{\ell\ell' \mu} p_{\ell' \ell \nu}}{(\Eell-\Eellp)^2} 
= -\sum_{\ell' \ne \ell} \int \Uell^\dagger \Uellpm d{\bm r} \int \Uellp^\dagger \Uelln d{\bm r} 
= \sum_{\ell'} \int \frac{\partial \Uell^\dagger}{\partial k_\mu} \Uellp d{\bm r} 
\int \Uellp^\dagger \Uelln d{\bm r} 
=\int \frac{\partial \Uell^\dagger}{\partial k_\mu}  
\frac{\partial \Uell}{\partial k_\nu}  d{\bm r}, 
\label{fSumRule2InA}
\end{equation}
where we have added the $\ell''=\ell$ term in (\ref{fSumRule3InA}) and 
$\ell' = \ell$ term in (\ref{fSumRule2InA}) which are equal to zero 
since $\int \Uell^\dagger (\partial \Uell/\partial k_\mu) dr =0$ (eq.~(\ref{AppBn2})). 

When we multiply $\Uell^\dagger$ to the $k_\mu, k_\nu, k_\tau$ derivative of eq.~(\ref{UellEq}) 
and make the real-space integral, we obtain
\begin{equation}
\frac{\partial^3 \Eell}{\partial k_\mu \partial k_\nu \partial k_\tau} 
= \int \Uell^\dagger \left( \frac{\partial H_{\bm k}}{\partial k_\mu} - \Eellm \right)
\frac{\partial^2 \Uell}{\partial k_\nu \partial k_\tau} d{\bm r} 
+\int \Uell^\dagger \left( \frac{\partial H_{\bm k}}{\partial k_\nu} - \Eelln \right)
\frac{\partial^2 \Uell}{\partial k_\tau \partial k_\mu} d{\bm r} 
+\int \Uell^\dagger \left( \frac{\partial H_{\bm k}}{\partial k_\tau} - \Eellt \right)
\frac{\partial^2 \Uell}{\partial k_\mu \partial k_\nu} d{\bm r},
\label{AppBe3_1}
\end{equation}
where we have used (\ref{AppBn2}). 
On the other hand, $k_\tau$ derivative of (\ref{AppB1}) gives
\begin{equation}
\int \Uell^\dagger  \frac{\partial H_{\bm k}}{\partial k_\mu} 
\frac{\partial^2 \Uell}{\partial k_\nu \partial k_\tau} d{\bm r} 
= -\int \frac{\partial \Uell^\dagger}{\partial k_\tau} \frac{\partial H_{\bm k}} {\partial k_\mu} \Uelln d{\bm r}
+\frac{1}{2}\frac{\partial^3 \Eell}{\partial k_\mu \partial k_\nu \partial k_\tau}.
\label{AppB1mod}
\end{equation}
Substituting this relation into the right-hand side of (\ref{AppBe3_1}) and using (\ref{AppBn3}), 
we can show
\begin{equation}
\frac{1}{2} \frac{\partial^3 \Eell}{\partial k_\mu \partial k_\nu \partial k_\tau} 
= \int \frac{\partial \Uell^\dagger}{\partial k_\tau} \left( \frac{\partial H_{\bm k}}{\partial k_\mu} - \Eellm \right)
\frac{\partial \Uell}{\partial k_\nu} d{\bm r} 
+\int \frac{\partial \Uell^\dagger}{\partial k_\mu} \left( \frac{\partial H_{\bm k}}{\partial k_\nu} - \Eelln \right)
\frac{\partial \Uell}{\partial k_\tau} d{\bm r} 
+\int \frac{\partial \Uell^\dagger}{\partial k_\nu} \left( \frac{\partial H_{\bm k}}{\partial k_\tau} - \Eellt \right)
\frac{\partial \Uell}{\partial k_\mu} d{\bm r}.
\label{AppBe3_2}
\end{equation}

\section{Calculations of $\chi_3$-$\chi_7$}

First, we calculate $\chi_3$, and $\chi_4$.
The summation over $n$ gives
\begin{equation}
\begin{split}
\chi_3 &= \frac{e^2}{\hbar^2 c^2} \sum_{\ell \ne \ell', {\bm k}} \Eellx  \Eellpx 
p_{\ell \ell' y}p_{\ell' \ell y} 
\left\{ \frac{f'(\Eell) +f'(\Eellp)}{(\Eell-\Eellp)^2} 
- 2\frac{f(\Eell) - f(\Eellp) }{(\Eell - \Eellp)^3}  \right\} + (x \leftrightarrow y) \cr 
&= \frac{2e^2}{\hbar^2 c^2} \sum_{\ell \ne \ell', {\bm k}} \Eellx  \Eellpx 
p_{\ell \ell' y}p_{\ell' \ell y} 
\left\{ \frac{f'(\Eell)}{(\Eell-\Eellp)^2} - \frac{2f(\Eell)}{(\Eell - \Eellp)^3}  \right\} 
+ (x \leftrightarrow y),
\label{AppC2}
\end{split}
\end{equation}
\begin{equation}
\begin{split}
\chi_4 &= \frac{2e^2}{\hbar^2 c^2} \sum_{\ell \ell' \ell'', {\bm k}}^{'}  \Eellx
p_{\ell \ell' y} p_{\ell' \ell'' x} p_{\ell'' \ell y}  
\biggl\{ \frac{f'(\Eell)}{(\Eell - \Eellp)(\Eell-\Eellpp)}  
- \frac{f(\Eell)}{(\Eell - \Eellp)^2(\Eell-\Eellpp)} + \frac{f(\Eellp)}{(\Eell - \Eellp)^2(\Eellp-\Eellpp)} \cr  
&\hskip 4.5truecm
- \frac{f(\Eell)}{(\Eell - \Eellp)(\Eell-\Eellpp)^2} + \frac{f(\Eellpp)}{(\Eell - \Eellpp)^2(\Eellpp-\Eellp)} 
\biggr\} + (x \leftrightarrow y). 
\label{AppC3}
\end{split}
\end{equation}
In the following, we denote the $j$-th term in the parentheses of $\chi_n$ as $\MyChi{n}{j}$. 
The summation over $\ell''$ in $\MyChi{4}{1}, \MyChi{4}{2}, \MyChi{4}{3}$ can be carried out
by using the formula (\ref{fSumRule3InA}) and its complex conjugate. 
As a result, we obtain
\begin{equation}
\MyChi{4}{1} = \frac{2e^2}{\hbar^2 c^2} \sum_{\ell \ne \ell', {\bm k}} \Eellx 
p_{\ell\ell' y} \frac{f'(\Eell)}{\Eell-\Eellp}  
\int \Uellp^\dagger \left( \frac{\partial H_{\bm k}}{\partial k_x} - \frac{\partial \Eellp}{\partial k_x}
\right) \Uelly d{\bm r} + (x \leftrightarrow y),
\label{AppC4}
\end{equation}
\begin{equation}
\MyChi{4}{2} = -\frac{2e^2}{\hbar^2 c^2} \sum_{\ell \ne \ell', {\bm k}} \Eellx 
p_{\ell\ell' y} \frac{f(\Eell)}{(\Eell-\Eellp)^2}  
\int \Uellp^\dagger \left( \frac{\partial H_{\bm k}}{\partial k_x} - \frac{\partial \Eellp}{\partial k_x}
\right) \Uelly d{\bm r} + (x \leftrightarrow y),
\end{equation}
and
\begin{equation}
\MyChi{4}{3} = \frac{2e^2}{\hbar^2 c^2} \sum_{\ell \ne \ell', {\bm k}} \Eellx 
p_{\ell \ell' y} \frac{f(\Eellp)}{(\Eell-\Eellp)^2}  
\int \frac{\partial \Uellp^\dagger}{\partial k_x} 
\left( \frac{\partial H_{\bm k}}{\partial k_y} - \frac{\partial \Eell}{\partial k_y}
\right) \Uell d{\bm r} + (x \leftrightarrow y).
\end{equation}
In the same way, the summation over $\ell'$ in $\MyChi{4}{4}$ and $\MyChi{4}{5}$ are 
carried out and we obtain
\begin{equation}
\MyChi{4}{4} = -\frac{2e^2}{\hbar^2 c^2} \sum_{\ell \ne \ell'', {\bm k}} \Eellx 
p_{\ell'' \ell y} \frac{f(\Eell)}{(\Eell-\Eellpp)^2}  
\int \frac{\partial \Uell^\dagger}{\partial k_y} 
\left( \frac{\partial H_{\bm k}}{\partial k_x} - \frac{\partial \Eellpp}{\partial k_x}
\right) \Uellpp d{\bm r} + (x \leftrightarrow y),
\end{equation}
\begin{equation}
\MyChi{4}{5} = \frac{2e^2}{\hbar^2 c^2} \sum_{\ell \ne \ell'', {\bm k}} \Eellx 
p_{\ell'' \ell y} \frac{f(\Eellpp)}{(\Eell-\Eellpp)^2}  
\int \Uell
\left( \frac{\partial H_{\bm k}}{\partial k_y} - \frac{\partial \Eell}{\partial k_y}
\right) \Uellppx d{\bm r} + (x \leftrightarrow y). 
\end{equation}

We find that the second term in $\MyChi{4}{1}$ cancels with $\MyChi{3}{1}$. 
For the first term in $\MyChi{4}{1}$, we can carry out the $\ell'$ summation 
by using the explicit form of $p_{\ell \ell' y}$ and the completeness property of $\Uellp$. 
As a result, we obtain 
\begin{equation}
\MyChi{3}{1} + \MyChi{4}{1} =\frac{2e^2}{\hbar^2 c^2} \sum_{\ell, {\bm k}} \Eellx f'(\Eell) 
\int \frac{\partial \Uell^\dagger}{\partial k_y} \frac{\partial H_{\bm k}}{\partial k_x}
\Uelly d{\bm r} + (x \leftrightarrow y).
\label{Chi3plus41}
\end{equation}
We also find that the second terms in $\MyChi{4}{2}$ and $\MyChi{4}{4}$ cancel with 
$\MyChi{3}{2}$. 
Then, rewriting $\MyChi{4}{4}$ and $\MyChi{4}{5}$
by $x \leftrightarrow y$, and making the change of variables $\Eellp\leftrightarrow \Eell$ etc., 
we obtain 
\begin{equation}
\begin{split}
&\MyChi{3}{2} + \MyChi{4}{2} + \MyChi{4}{3} + \MyChi{4}{4} + \MyChi{4}{5}\cr
&=\frac{2e^2}{\hbar^2 c^2} \sum_{\ell \ne \ell', {\bm k}} 
\frac{f(\Eell)}{\Eell-\Eellp} \biggl\{ 
\Eellx \int \Uell^\dagger \Uellpy d{\bm r} 
\int \Uellp^\dagger \frac{\partial H_{\bm k}}{\partial k_x} \Uelly d{\bm r} 
-\Eelly \int \frac{\partial \Uell^\dagger}{\partial k_x}  
\frac{\partial H_{\bm k}}{\partial k_y} \Uellp d{\bm r}
\int \Uellp^\dagger \Uellx d{\bm r} \cr
&\qquad + \Eellpx 
\int \frac{\partial \Uell^\dagger}{\partial k_x} 
\left( \frac{\partial H_{\bm k}}{\partial k_y} - \frac{\partial \Eellp}{\partial k_y}
\right) \Uellp d{\bm r} 
\int \Uellp \Uelly d{\bm r}
- \frac{\partial \Eellp}{\partial k_y}
\int \Uell \Uellpx 
\int \Uellp
\left( \frac{\partial H_{\bm k}}{\partial k_x} - \frac{\partial \Eellp}{\partial k_x}
\right) \Uelly d{\bm r} \biggr\}  + (x \leftrightarrow y),
\label{AppC6}
\end{split}
\end{equation}

Next, we calculate $\chi_5$, $\chi_6$, and $\chi_7$.
The summation over $n$ gives
\begin{equation}
\chi_5 = \frac{e^2}{\hbar^2 c^2} \sum_{\ell \ne \ell', {\bm k}} 
p_{\ell\ell' x} p_{\ell' \ell y} p_{\ell\ell' x} p_{\ell' \ell y} 
\left\{ \frac{f'(\Eell)}{(\Eell - \Eellp)^2} - \frac{2f(\Eell)}{(\Eell - \Eellp)^3} \right\} 
+ (x \leftrightarrow y),
\label{AppD1}
\end{equation}
\begin{equation}
\chi_6 = \frac{e^2}{\hbar^2 c^2} \sum_{\ell \ell' \ell'', {\bm k}}^{'} 
p_{\ell\ell' x} p_{\ell' \ell y} p_{\ell\ell'' x} p_{\ell'' \ell y} 
\left\{ \frac{f'(\Eell)}{(\Eell - \Eellp)(\Eell - \Eellpp)}  
- \frac{2f(\Eell)}{(\Eell - \Eellp)^2 (\Eell - \Eellpp)} 
+ \frac{2f(\Eellp)}{(\Eell - \Eellp)^2 (\Eellp - \Eellpp)} \right\} + (x \leftrightarrow y),
\label{AppD2}
\end{equation}
\begin{equation}
\chi_7 = \frac{2e^2}{\hbar^2 c^2} \sum_{\ell \ell' \ell'' \ell''', {\bm k}}^{'}
p_{\ell\ell' x} p_{\ell' \ell'' y} p_{\ell'' \ell''' x} p_{\ell''' \ell y} 
\frac{f(\Eell)}{(\Eell - \Eellp)(\Eell - \Eellpp)(\Eell - \varepsilon_{\ell'''})}  + (x \leftrightarrow y),
\label{AppD3}
\end{equation}
where we have arranged the terms by using change of variables, $\ell\leftrightarrow \ell'$, etc. 
We can see that the excluded terms of $\ell'=\ell''$ in 
$\MyChi{6}{1}$ and $\MyChi{6}{2}$ are exactly supplemented by 
$\MyChi{5}{1}$ and $\MyChi{5}{2}$, respectively. 
Therefore the $\ell'$- and $\ell''$-summations in $\MyChi{5}{1} + \MyChi{6}{1}$ and 
$\MyChi{5}{2} + \MyChi{6}{2}$ are independent to each other, and we can use 
the $f$-sum rule (\ref{fSumRuleInA}) and (\ref{fSumRule2InA}).  
As a result, we obtain
\begin{equation}
\MyChi{5}{1} + \MyChi{6}{1} = \frac{e^2}{\hbar^2 c^2} \sum_{\ell, {\bm k}} f'(\Eell)
\sum_{\ell'} \frac{p_{\ell\ell' x} p_{\ell' \ell y}}{\Eell-\Eellp}  
\sum_{\ell''} \frac{p_{\ell\ell'' x} p_{\ell'' \ell y}}{\Eell-\Eellpp} + (x \leftrightarrow y)
= \frac{e^2}{4\hbar^2 c^2} \sum_{\ell, {\bm k}} 
f'(\Eell) \left( \frac{\partial^2 \Eell}{\partial k_x \partial k_y} \right)^2 + (x \leftrightarrow y),
\label{Chi5plus61}
\end{equation}
which is shown in (\ref{Chi56X}), and
\begin{equation}
\MyChi{5}{2} + \MyChi{6}{2} = -\frac{e^2}{\hbar^2 c^2} \sum_{\ell, {\bm k}} 
f(\Eell) \frac{\partial^2 \Eell}{\partial k_x \partial k_y}
\int \frac{\partial \Uell^\dagger}{\partial k_x} \Uelly d{\bm r} + (x \leftrightarrow y).
\label{Chi5plus62}
\end{equation}

Similarly, we can see that the excluded terms of $\ell' = \ell'''$ in $\chi_7$ (eq.~(\ref{AppD3})) 
are exactly supplemented by $\MyChi{6}{3}$ in (\ref{AppD2}) 
by changing $\ell\leftrightarrow \ell'$ and $x\leftrightarrow y$ in $\MyChi{6}{3}$.  
As a result, applying (\ref{fSumRule3InA}) and its complex conjugates to 
$\ell'$- and $\ell'''$-summations, we obtain
\begin{equation}
\begin{split}
\MyChi{6}{3} +\chi_7 &=  \frac{2e^2}{\hbar^2 c^2} \sum_{\ell \ne \ell'', {\bm k}} \frac{f(\Eell)}{(\Eell - \Eellpp)} 
\int \frac{\partial \Uell^\dagger}{\partial k_x} \left(\frac{\partial H_{\bm k}}{\partial k_y} - 
\frac{\partial \Eellpp}{\partial k_y} \right) \Uellpp d{\bm r}  
\int \Uellpp \left(\frac{\partial H_{\bm k}}{\partial k_x} - \frac{\partial \Eellpp}{\partial k_x} \right)
\Uelly d{\bm r} + (x \leftrightarrow y).
\label{AppD5}
\end{split}
\end{equation}

Now all the contributions are arranged into the forms which have either $f(\Eell)$, $f'(\Eell)$, 
or $f(\Eell)/(\Eell-\Eellp)$. 
The contributions with $f(\Eell)/(\Eell-\Eellp)$ are $\MyChi{2}{3}$ in (\ref{Chi2-3}), 
$\MyChi{3}{2} + \MyChi{4}{2} + \MyChi{4}{3} + \MyChi{4}{4} + \MyChi{4}{5}$ 
in (\ref{AppC6}) and $\MyChi{6}{3} +\chi_7$ in (\ref{AppD5}). 
Comparing each term and using the $x\leftrightarrow y$ terms, 
we can see that the several terms cancel with each other and 
the remaining terms can be rewritten in a compact form as
\begin{equation}
\begin{split}
&\MyChi{2}{3} + \MyChi{3}{2} + \MyChi{4}{2} + \MyChi{4}{3} + \MyChi{4}{4} + \MyChi{4}{5}
+\MyChi{6}{3} +\chi_7  \cr
&=\frac{2e^2}{\hbar^2 c^2} \sum_{\ell \ne \ell', {\bm k} } \frac{f(\Eell)}{\Eell - \Eellp} \left( 
\int \frac{\partial \Uell^\dagger}{\partial k_x} \frac{\partial H_{\bm k}}{\partial k_y} \Uellp d{\bm r}  
-\frac{\partial \Eell}{\partial k_x} \int \frac{\partial \Uell^\dagger}{\partial k_y} \Uellp d{\bm r} \right)
\left( 
\int \Uellp^\dagger \frac{\partial H_{\bm k}}{\partial k_x} \Uelly d{\bm r} 
-\frac{\partial \Eell }{\partial k_y} \int \Uellp^\dagger \Uellx d{\bm r} \right) + (x \leftrightarrow y).
\label{chitotalinter0}
\end{split}
\end{equation}

We find it more convenient to rewrite (\ref{chitotalinter0}) as follows:
First from eq.~(\ref{AppB3p}), we obtain for $\ell\ne \ell'$
\begin{equation}
   \int \Uellp^\dagger \frac{\partial H_{\bm k}}{\partial k_x} \Uelly d{\bm r} 
- \int \Uellp^\dagger \frac{\partial \Eell}{\partial k_y} \Uellx d{\bm r} 
= - \int \Uellp^\dagger \frac{\partial H_{\bm k}}{\partial k_y} \Uellx d{\bm r} 
+   \int \Uellp^\dagger \frac{\partial \Eell}{\partial k_x} \Uelly d{\bm r} 
+ \left( \Eell - \Eellp \right) \int \Uellp^\dagger \frac{\partial^2 \Uell}{\partial k_x \partial k_y}
d{\bm r}.
\label{AppD11}
\end{equation}
Using this relation, we can show 
\begin{equation}
\begin{split}
   \int \Uellp^\dagger \frac{\partial H_{\bm k}}{\partial k_x} \Uelly d{\bm r} 
- \int \Uellp^\dagger \frac{\partial \Eell}{\partial k_y} \Uellx d{\bm r} 
&= \frac{1}{2} \biggl\{
\int \Uellp^\dagger \left( \frac{\partial H_{\bm k}}{\partial k_x} + \frac{\partial \Eell}{\partial k_x}
\right) \Uelly d{\bm r} -
\int \Uellp^\dagger \left( \frac{\partial H_{\bm k}}{\partial k_y} + \frac{\partial \Eell}{\partial k_y}
\right) \Uellx d{\bm r} \cr
&\qquad 
+ \left( \Eell - \Eellp \right) \int \Uellp^\dagger \frac{\partial^2 \Uell}{\partial k_x \partial k_y}
d{\bm r}\biggr\}.
\label{AppD12}
\end{split}
\end{equation}
With use of this formula, (\ref{chitotalinter0}) can be rewritten as 
\begin{equation}
\begin{split}
& \frac{e^2}{2\hbar^2 c^2} \sum_{\ell \ne \ell', {\bm k} } \frac{f(\Eell)}{\Eell - \Eellp} 
\biggl\{
\int \frac{\partial \Uell^\dagger}{\partial k_x} 
\left( \frac{\partial H_{\bm k}}{\partial k_y} + \frac{\partial \Eell}{\partial k_y} \right) \Uellp d{\bm r} 
- \int \frac{\partial \Uell^\dagger}{\partial k_y} 
\left( \frac{\partial H_{\bm k}}{\partial k_x} + \frac{\partial \Eell}{\partial k_x} \right) \Uellp  d{\bm r} 
+ \left( \Eell - \Eellp \right) \int \frac{\partial^2 \Uell^\dagger}{\partial k_x \partial k_y}
\Uellp d{\bm r} \biggr\} \cr
&\qquad \quad \times \biggl\{
\int \Uellp^\dagger \left( \frac{\partial H_{\bm k}}{\partial k_x} + \frac{\partial \Eell}{\partial k_x}
\right) \Uelly d{\bm r} -
\int \Uellp^\dagger \left( \frac{\partial H_{\bm k}}{\partial k_y} + \frac{\partial \Eell}{\partial k_y}
\right) \Uellx d{\bm r} 
+ \left( \Eell - \Eellp \right) \int \Uellp^\dagger \frac{\partial^2 \Uell}{\partial k_x \partial k_y}
d{\bm r}\biggr\}  + (x \leftrightarrow y)  \cr
&= -\frac{e^2}{2\hbar^2 c^2} \sum_{\ell \ne \ell', {\bm k}} \frac{f(\Eell)}{\Eell - \Eellp} 
\biggl|
\int \frac{\partial \Uell^\dagger}{\partial k_x} 
\left( \frac{\partial H_{\bm k}}{\partial k_y} + \frac{\partial \Eell}{\partial k_y} \right) \Uellp d{\bm r} 
- \int \frac{\partial \Uell^\dagger}{\partial k_y} 
\left( \frac{\partial H_{\bm k}}{\partial k_x} + \frac{\partial \Eell}{\partial k_x} \right) \Uellp 
d{\bm r} \biggr|^2 + (x \leftrightarrow y) \cr 
&+ \frac{e^2}{2\hbar^2 c^2} \sum_{\ell \ne \ell', {\bm k}}
f(\Eell) (\Eell - \Eellp) \int \frac{\partial^2 \Uell^\dagger}{\partial k_x \partial k_y}\Uellp d{\bm r}
\int \Uellp^\dagger \frac{\partial^2 \Uell}{\partial k_x \partial k_y} d{\bm r} + (x \leftrightarrow y),
\label{AppD13}
\end{split}
\end{equation}
where the cross terms vanish. 
We can rewrite the second summation in (\ref{AppD13}) by using the 
completeness property of $\Uellp$ and the complex conjugate of eq.~(\ref{AppBx}) as
\begin{equation}
\begin{split}
\frac{e^2}{2\hbar^2 c^2} \sum_{\ell, {\bm k}} f(\Eell) 
\biggl\{ 2\int \frac{\partial \Uell^\dagger}{\partial k_y} \left( 
\frac{\partial H_{\bm k}}{\partial k_x} - \frac{\partial \Eell}{\partial k_x} \right) 
\frac{\partial^2\Uell}{\partial k_x \partial k_y} d{\bm r} 
+ \frac{\partial^2 \Eell}{\partial k_x \partial k_y} \int \frac{\partial \Uell^\dagger}{\partial k_x}  
\frac{\partial \Uell}{\partial k_y} d{\bm r}\biggr\} + (x \leftrightarrow y),
\label{AppD14}
\end{split}
\end{equation}
where we have used the relation (\ref{AppBn3}). 

Here we collect all the contributions as follows:
\begin{equation}
\begin{split}
\chi_1 + \MyChi{2}{1} + \MyChi{5}{1} + \MyChi{6}{1} &= \frac{e^2}{6\hbar^2 c^2} \sum_{\ell, {\bm k}} f'(\Eell) 
\biggl\{ \frac{\partial^2 \Eell}{\partial k_x^2} \frac{\partial^2 \Eell}{\partial k_y^2} 
-\left( \frac{\partial^2 \Eell}{\partial k_x \partial k_y} \right)^2 
-\frac{3}{2} \left( 
\Eellx \frac{\partial^3 \varepsilon_\ell}{\partial k_x \partial k_y^2} + 
\Eelly \frac{\partial^3 \varepsilon_\ell}{\partial k_x^2 \partial k_y}
\right) \biggr\}, 
\label{ChiTotal1to6}
\end{split}
\end{equation}
\begin{equation}
\begin{split}
\MyChi{2}{2} + \MyChi{3}{1} + \MyChi{4}{1} &=\frac{2e^2}{\hbar^2 c^2} \sum_{\ell, {\bm k}} f'(\Eell) \biggl\{ 
\Eellx\int \frac{\partial \Uell^\dagger}{\partial k_y} 
\frac{\partial H_{\bm k}}{\partial k_x}  \Uelly d{\bm r}  
-\Eellx\Eelly \int \frac{\partial \Uell^\dagger}{\partial k_x} 
\Uelly d{\bm r} \biggr\} + (x\leftrightarrow y), 
\label{ChiTotal2to4}
\end{split}
\end{equation}
\begin{equation}
\begin{split}
(\MyChi{2}{3} + \MyChi{3}{2} + &\MyChi{4}{2} + \MyChi{4}{3} + \MyChi{4}{4} + \MyChi{4}{5}
+\MyChi{6}{3} +\chi_7) + ( \MyChi{5}{2} + \MyChi{6}{2})  \cr
&= -\frac{e^2}{\hbar^2 c^2} \sum_{\ell \ne \ell', {\bm k} } \frac{f(\Eell)}{\Eell - \Eellp} 
\biggl| \int \frac{\partial \Uell^\dagger}{\partial k_x} 
\left( \frac{\partial H_{\bm k}}{\partial k_y} + \frac{\partial \Eell}{\partial k_y} \right) \Uellp d{\bm r} 
- \int \frac{\partial \Uell^\dagger}{\partial k_y} 
\left( \frac{\partial H_{\bm k}}{\partial k_x} + \frac{\partial \Eell}{\partial k_x} \right) \Uellp 
d{\bm r} \biggr|^2 \cr
&\quad + \frac{e^2}{2\hbar^2 c^2} \sum_{\ell, {\bm k}} f(\Eell) 
\biggl\{ 2\int \frac{\partial \Uell^\dagger}{\partial k_y} \left( 
\frac{\partial H_{\bm k}}{\partial k_x} - \frac{\partial \Eell}{\partial k_x} \right) 
\frac{\partial^2\Uell}{\partial k_x \partial k_y} d{\bm r} 
- \frac{\partial^2 \Eell}{\partial k_x \partial k_y} \int \frac{\partial \Uell^\dagger}{\partial k_x}  
\frac{\partial \Uell}{\partial k_y} d{\bm r} \biggr\} + (x \leftrightarrow y),
\label{ChiTotal1}
\end{split}
\end{equation}
Here the first two terms in (\ref{ChiTotal1to6}) are just equal to the 
Landau-Peierls susceptibility, $\chi_{\rm LP}$. 
The first summation in (\ref{ChiTotal1}) involving two bands ($\ell$ and $\ell'$) 
is denoted as $\chi_{\rm inter}$. 
In the following, we rewrite the remaining terms a little further. 

As discussed in Section 2, the last term in (\ref{ChiTotal1to6}) can be rewritten by 
using the formula in eq.~(\ref{E3Formula}). 
As a result, sum of  (\ref{ChiTotal1to6}) and  (\ref{ChiTotal2to4}) becomes
\begin{equation}
\begin{split}
\chi_{\rm LP} +\frac{e^2}{\hbar^2 c^2} \sum_{\ell, {\bm k}} f'(\Eell) \biggl\{ 
\Eellx\int \frac{\partial \Uell^\dagger}{\partial k_y} 
\left( \frac{3}{2} \frac{\partial H_{\bm k}}{\partial k_x} + 
\frac{1}{2} \frac{\partial \Eell}{\partial k_x} \right)
\Uelly d{\bm r} 
-\Eellx\int \frac{\partial \Uell^\dagger}{\partial k_x} 
\left( \frac{\partial H_{\bm k}}{\partial k_y} + \frac{\partial \Eell}{\partial k_y} \right)
\Uelly d{\bm r}\biggr\} + (x\leftrightarrow y). 
\label{ChiIntraPre}
\end{split}
\end{equation}
Furthermore, we notice that it is convenient to rewrite the first term of 
the second summation in (\ref{ChiTotal1}) as
\begin{equation}
\begin{split}
&\frac{e^2}{2\hbar^2 c^2} \sum_{\ell, {\bm k}} f(\Eell) \biggl\{
2\int \frac{\partial \Uell^\dagger}{\partial k_y} \left( 
\frac{\partial H_{\bm k}}{\partial k_x} - \frac{\partial \Eell}{\partial k_x} \right) 
\frac{\partial^2\Uell}{\partial k_x \partial k_y} d{\bm r}\biggr\}  \cr
=&\frac{e^2}{2\hbar^2 c^2} \sum_{\ell, {\bm k}} f(\Eell) \biggl\{
\frac{\partial}{\partial k_x} \left[ \int \frac{\partial \Uell^\dagger}{\partial k_y} \left( 
\frac{\partial H_{\bm k}}{\partial k_x} - \frac{\partial \Eell}{\partial k_x} \right) 
\frac{\partial \Uell}{\partial k_y} d{\bm r}\right] 
-\left( \frac{\hbar^2}{m} - \frac{\partial^2 \Eell}{\partial k_x^2} \right)
\int \frac{\partial \Uell^\dagger}{\partial k_y}  
\frac{\partial \Uell}{\partial k_y} d{\bm r} \biggr\} \cr
=&-\frac{e^2}{2\hbar^2 c^2} \sum_{\ell, {\bm k}} f'(\Eell) 
\Eellx \int \frac{\partial \Uell^\dagger}{\partial k_y} \left( 
\frac{\partial H_{\bm k}}{\partial k_x} - \frac{\partial \Eell}{\partial k_x} \right) 
\frac{\partial \Uell}{\partial k_y} d{\bm r} 
-\frac{e^2}{2\hbar^2 c^2} \sum_{\ell, {\bm k}} f(\Eell) 
\left( \frac{\hbar^2}{m} - \frac{\partial^2 \Eell}{\partial k_x^2} \right)
\int \frac{\partial \Uell^\dagger}{\partial k_y}  
\frac{\partial \Uell}{\partial k_y} d{\bm r},
\label{ChiInterPre}
\end{split}
\end{equation}
where we have used the relation (\ref{AppBn4}).
It is now apparent that the summation in (\ref{ChiIntraPre}) 
and (\ref{ChiInterPre}) should be added.  
Using these expressions, we finally obtain 
$\chi=\chi_{\rm LP}+\chi_{\rm inter}+ \chi_{\rm FS} + \chi_{\rm occ}$ as shown in (\ref{FinalChi}).

\section{Proof of equivalence of the present result with that obtained by HLSS}

In this Appendix, we prove the equivalence of the present result with HLSS's result; 
eqs.~(4.3)-(4.6) in Ref.\cite{HS3} (or equivalently eqs.~(53)-(56) in Ref.\cite{HS2}). 
In the notation of the present paper, their result is given by
\begin{equation}
\begin{split}
\chi^{({\rm HLSS})} &= \chi_1^{({\rm HLSS})} + \chi_2^{({\rm HLSS})} 
+ \chi_3^{({\rm HLSS})} + \chi_4^{({\rm HLSS})}, \cr
\chi_1^{({\rm HLSS})} &= \frac{e^2}{6\hbar^2 c^2} 
\sum_{\ell, {\bm k}} f'(\Eell) 
\biggl\{ \frac{\partial^2 \Eell}{\partial k_x^2} \frac{\partial^2 \Eell}{\partial k_y^2} 
-\left( \frac{\partial^2 \Eell}{\partial k_x \partial k_y} \right)^2
+\frac{3}{2} \left( \Eellx \frac{\partial^3 \Eell}{\partial k_x \partial k_y^2} 
+ \Eelly \frac{\partial^3 \Eell}{\partial k_x^2 \partial k_y} \right) \biggr\}, \cr
\chi_2^{({\rm HLSS})} &= \frac{2e^2}{\hbar^2 c^2}\sum_{\ell, {\bm k}} f'(\Eell) \Eelly 
\biggl\{ 
\Eelly \int \frac{\partial \Uell^\dagger}{\partial k_x} \Uellx d{\bm r} 
-\int \frac{\partial \Uell^\dagger}{\partial k_x} \frac{\partial H_{\bm k}}{\partial k_x} \Uelly d{\bm r} 
-\int \frac{\partial \Uell^\dagger}{\partial k_y} \frac{\partial H_{\bm k}}{\partial k_x} \Uellx d{\bm r}
\biggr\}, \cr
\chi_3^{({\rm HLSS})} &= - \frac{4e^2}{\hbar^2 c^2} \sum_{\ell \ne \ell', {\bm k} } 
\frac{f(\Eell)}{(\Eell - \Eellp)} \biggl\{ 
\int \frac{\partial \Uell^\dagger}{\partial k_x} \frac{\partial H_{\bm k}}{\partial k_y} \Uellp d{\bm r} 
-\int \frac{\partial \Uell^\dagger}{\partial k_y} \frac{\partial H_{\bm k}}{\partial k_x} \Uellp d{\bm r}
+ \Eellx \int \Uell^\dagger \frac{\partial \Uellp}{\partial k_y} d{\bm r} \biggr\} 
\int \Uellp^\dagger \Uellx d{\bm r} \Eelly, \cr
\chi_4^{({\rm HLSS})} &= -  \frac{2e^2}{\hbar^2 c^2} \sum_{\ell, {\bm k}} \frac{\hbar^2}{m} 
f(\Eell) \int \frac{\partial \Uell^\dagger}{\partial k_y} \Uelly d{\bm r} 
- \frac{4e^2}{\hbar^2 c^2}\sum_{\ell \ne \ell', {\bm k} } \frac{f(\Eell)}{(\Eell - \Eellp)} 
\biggl| \int \frac{\partial \Uell^\dagger}{\partial k_y} 
\frac{\partial H_{\bm k}}{\partial k_x} \Uellp d{\bm r} \biggr|^2. 
\label{HSChi}
\end{split}
\end{equation}

First, using the complex conjugate of (\ref{AppD11}), we rewrite the last term of 
$\chi_4^{({\rm HLSS})}$ as 
\begin{equation}
\begin{split}
-\frac{e^2}{\hbar^2 c^2}\sum_{\ell \ne \ell', {\bm k} } \frac{f(\Eell)}{(\Eell - \Eellp)} &\biggl| 
   \int \frac{\partial \Uell^\dagger}{\partial k_y} \frac{\partial H_{\bm k}}{\partial k_x} \Uellp d{\bm r} 
- \int \frac{\partial \Uell^\dagger}{\partial k_x} \frac{\partial H_{\bm k}}{\partial k_y} \Uellp d{\bm r} 
+ \int \frac{\partial \Uell^\dagger}{\partial k_y} \frac{\partial \Eell}{\partial k_x} \Uellp d{\bm r} 
+ \int \frac{\partial \Uell^\dagger}{\partial k_x} \frac{\partial \Eell}{\partial k_y} \Uellp d{\bm r} \cr
&+ \left( \Eell - \Eellp \right) \int \frac{\partial^2 \Uell^\dagger}{\partial k_x \partial k_y} \Uellp d{\bm r}
\biggr|^2. 
\end{split}
\end{equation}
Then combining with $\chi_3^{({\rm HLSS})}$, we obtain
\begin{equation}
\begin{split}
\chi_3^{({\rm HLSS})} + &{\rm last}\ {\rm term}\ {\rm of} \ \chi_4^{({\rm HLSS})} = 
-\frac{e^2}{\hbar^2 c^2}\sum_{\ell \ne \ell', {\bm k} } \frac{f(\Eell)}{(\Eell - \Eellp)} 
\biggl[ \biggl| 
   \int \frac{\partial \Uell^\dagger}{\partial k_y} 
\left( \frac{\partial H_{\bm k}}{\partial k_x} + \Eellx \right) \Uellp d{\bm r} 
- \int \frac{\partial \Uell^\dagger}{\partial k_x} 
\left( \frac{\partial H_{\bm k}}{\partial k_y} + \Eelly \right) \Uellp d{\bm r} 
\biggr|^2 \cr
& \qquad \qquad + 2(\Eell-\Eellp) {\rm Re} \left\{ 
 \int \frac{\partial \Uell^\dagger}{\partial k_y} 
\left( \frac{\partial H_{\bm k}}{\partial k_x} + \Eellx \right) \Uellp d{\bm r} 
- \int \frac{\partial \Uell^\dagger}{\partial k_x}
\left( \frac{\partial H_{\bm k}}{\partial k_y} - \Eelly \right) \Uellp d{\bm r} \right\}
\int \Uellp^\dagger \frac{\partial^2 \Uell}{\partial k_x \partial k_y} d{\bm r} \cr
& \qquad \qquad + (\Eell-\Eellp)^2 
\int \frac{\partial^2 \Uell^\dagger}{\partial k_x \partial k_y} \Uellp d{\bm r}
\int \Uellp^\dagger \frac{\partial^2 \Uell}{\partial k_x \partial k_y} d{\bm r} \biggr] \cr 
&  \qquad \quad
= -\frac{e^2}{\hbar^2 c^2}\sum_{\ell \ne \ell', {\bm k} } \frac{f(\Eell)}{(\Eell - \Eellp)} \biggl| 
   \int \frac{\partial \Uell^\dagger}{\partial k_y} 
\left( \frac{\partial H_{\bm k}}{\partial k_x} + \Eellx \right) \Uellp d{\bm r} 
- \int \frac{\partial \Uell^\dagger}{\partial k_x} 
\left( \frac{\partial H_{\bm k}}{\partial k_y} + \Eelly \right) \Uellp d{\bm r} 
\biggr|^2 \cr
& \qquad \qquad -\frac{e^2}{\hbar^2 c^2}\sum_{\ell, {\bm k} } f(\Eell) \biggl\{ 
 2\int \frac{\partial \Uell^\dagger}{\partial k_y} 
\left( \frac{\partial H_{\bm k}}{\partial k_x} + \Eellx \right) 
\frac{\partial^2 \Uell}{\partial k_x \partial k_y} d{\bm r} 
- 2\int \frac{\partial \Uell^\dagger}{\partial k_x}
\left( \frac{\partial H_{\bm k}}{\partial k_y} - \Eelly \right) 
\frac{\partial^2 \Uell}{\partial k_x \partial k_y} d{\bm r} \cr 
& \qquad \qquad \qquad \qquad \qquad \qquad 
+ \int \frac{\partial^2 \Uell^\dagger}{\partial k_x \partial k_y} \left(\Eell - H_{\bm k} \right)  
\frac{\partial^2 \Uell}{\partial k_x \partial k_y} d{\bm r} \biggr\},
\end{split}
\end{equation}
where we have used the completeness property of $\Uellp$. 
Now the first term on the right-hand side is the same as $\chi_{\rm inter}$ of eq.~(\ref{ChiInter}).
The second summation can be rewritten by using the complex conjugate of eq.~(\ref{AppBx}) as
\begin{equation}
\begin{split}
&-\frac{e^2}{\hbar^2 c^2}\sum_{\ell, {\bm k} } f(\Eell) \biggl\{ 
 \int \frac{\partial \Uell^\dagger}{\partial k_y} 
\left( 3\frac{\partial H_{\bm k}}{\partial k_x} + \Eellx \right) 
\frac{\partial^2 \Uell}{\partial k_x \partial k_y} d{\bm r} 
- \int \frac{\partial \Uell^\dagger}{\partial k_x}
\left( \frac{\partial H_{\bm k}}{\partial k_y} - \Eelly \right) 
\frac{\partial^2 \Uell}{\partial k_x \partial k_y} d{\bm r} 
+ \frac{\partial^2 \Eell}{\partial k_x \partial k_y}
\int \frac{\partial \Uell^\dagger}{\partial k_x}
\frac{\partial \Uell}{\partial k_y} d{\bm r} \biggr\} \cr
=&-\frac{e^2}{\hbar^2 c^2}\sum_{\ell, {\bm k} } f(\Eell) \biggl\{ 
\frac{1}{2} \frac{\partial }{\partial k_x} \left[ \int \frac{\partial \Uell^\dagger}{\partial k_y} 
\left( 3\frac{\partial H_{\bm k}}{\partial k_x} + \Eellx \right) 
\frac{\partial \Uell}{\partial k_y} d{\bm r} \right]
- \frac{1}{2}\frac{\partial }{\partial k_y} \left[ \int \frac{\partial \Uell^\dagger}{\partial k_x}
\left( \frac{\partial H_{\bm k}}{\partial k_y} - \Eelly \right) 
\frac{\partial \Uell}{\partial k_x} d{\bm r} \right] \cr
&\qquad\qquad \qquad 
- \left( \frac{3}{2} \frac{\hbar^2}{m} + \frac{1}{2} \frac{\partial^2 \Eell}{\partial k_x^2} \right) 
\int \frac{\partial \Uell^\dagger}{\partial k_y} \frac{\partial \Uell}{\partial k_y} d{\bm r}
+ \left( \frac{1}{2} \frac{\hbar^2}{m} - \frac{1}{2} \frac{\partial^2 \Eell}{\partial k_y^2} \right) 
\int \frac{\partial \Uell^\dagger}{\partial k_x} \frac{\partial \Uell}{\partial k_x} d{\bm r}
+ \frac{\partial^2 \Eell}{\partial k_x \partial k_y}
\int \frac{\partial \Uell^\dagger}{\partial k_x}
\frac{\partial \Uell}{\partial k_y} d{\bm r} \biggr\},
\label{AppE4new}
\end{split}
\end{equation}
where we have used (\ref{AppBn3}) and (\ref{AppBn4}).
We find that the last three terms plus the first term of $\chi_4^{({\rm HLSS})}$
are equal to $\chi_{\rm occ}$ in (\ref{ChiC}). 

Using these relations and integration by parts, we can now write 
\begin{equation}
\begin{split}
\chi^{({\rm HLSS})} &= \chi_{\rm LP}+\chi_{\rm inter}+\chi_{\rm occ} 
+ \frac{e^2}{4\hbar^2 c^2} \sum_{\ell, {\bm k}} f'(\Eell) 
\left( \Eellx \frac{\partial^3 \Eell}{\partial k_x \partial k_y^2} 
+ \Eelly \frac{\partial^3 \Eell}{\partial k_x^2 \partial k_y} \right) + \chi_2^{({\rm HLSS})} \cr 
&+\frac{e^2}{\hbar^2 c^2}\sum_{\ell, {\bm k} } f'(\Eell) \biggl\{ 
\Eellx \int \frac{\partial \Uell^\dagger}{\partial k_y} 
\left( \frac{3}{2} \frac{\partial H_{\bm k}}{\partial k_x} + \frac{1}{2} \Eellx \right) 
\frac{\partial \Uell}{\partial k_y} d{\bm r} 
- \Eelly \int \frac{\partial \Uell^\dagger}{\partial k_x}
\left( \frac{1}{2} \frac{\partial H_{\bm k}}{\partial k_y} - \frac{1}{2} \Eelly \right) 
\frac{\partial \Uell}{\partial k_x} d{\bm r} \biggr\}. 
\label{AppE4}
\end{split}
\end{equation}
The fourth term can be rewritten in terms of $\Uell$ as carried out in 
eq.~(\ref{E3Formula}). 
Then collecting all the terms, we obtain
\begin{equation}
\begin{split}
\chi^{({\rm HLSS})} = \chi_{\rm LP}+\chi_{\rm inter}+\chi_{\rm occ} 
+\frac{e^2}{\hbar^2 c^2}\sum_{\ell, {\bm k} } f'(\Eell) \biggl\{ 
&2\Eellx \int \frac{\partial \Uell^\dagger}{\partial k_y} 
\frac{\partial H_{\bm k}}{\partial k_x}  \frac{\partial \Uell}{\partial k_y} d{\bm r} 
+ \Eellx \int \frac{\partial \Uell^\dagger}{\partial k_x}
\left( \frac{\partial H_{\bm k}}{\partial k_y} - \Eelly \right) 
\frac{\partial \Uell}{\partial k_y} d{\bm r} \cr
+&2\Eelly \int \frac{\partial \Uell^\dagger}{\partial k_x} 
\frac{\partial \Eell}{\partial k_y}  \frac{\partial \Uell}{\partial k_x} d{\bm r} 
- \Eelly \int \frac{\partial \Uell^\dagger}{\partial k_y}
\left( 3\frac{\partial H_{\bm k}}{\partial k_x} + \Eelly \right) 
\frac{\partial \Uell}{\partial k_x} d{\bm r} \biggr\}.
\label{AppE5}
\end{split}
\end{equation}
Apparently (\ref{AppE5}) is not symmetric with respect to the exchange of $x$ and $y$. 
This is because HLSS used the Landau gauge ${\bm A}=(-yH, 0, 0)$. 
As mentioned by Hebborn and Sondheimer\cite{HS2}
if the other gauge ${\bm A}=(0, xH, 0)$ is chosen, a result in which $x$ and $y$ interchanged 
will be obtained. 
Thus by taking the average of this expression and the original one, we can have a 
result that is symmetric with respect to $x$ and $y$. 
After this symmetrization, we can see that (\ref{AppE5}) is equal to
$\chi = \chi_{\rm LP} + \chi_{\rm inter} + \chi_{\rm FS} + \chi_{\rm occ}$. 

\section{Comparison of the present result with that obtained by Gao {\it et al}}

In this Appendix, we compare our results with those obtained by Gao {\it et al}\cite{Gao} 
in the case where the time-reversal symmetry is not broken and the Berry curvature 
$\Omega_z$ is equal to zero. 
In this case, the thermodynamic potential per spin 
in the second order of $\bm B$ becomes (see eq.~(6) in Ref.\cite{Gao})
\begin{equation}
\begin{split}
\frac{\hbar^2 c^2}{2e^2} \Omega^{(2)} &= g_L - 
\sum_{\ell, {\bm k}} \frac{f'(\Eell)}{4} {\bm v}_0 \cdot {\bm P}_{\rm E} 
+\sum_{\ell \ne \ell', {\bm k}} \frac{f(\Eell)}{\Eell - \Eellp} G_{\ell\ell'}G_{\ell'\ell}
+\sum_{\ell, {\bm k}} \frac{f(\Eell)}{8m}(B^2 g_{ii}-B_i g_{ij}B_j ) 
-\sum_{\ell, {\bm k}} \frac{f(\Eell)}{8} \varepsilon_{ijk} \varepsilon_{i'j'k'} 
B_i B_{i'} g_{jj'} \alpha_{kk'},
\label{GaoChi}
\end{split}
\end{equation}
where $g_L$ represents the Landau-Peierls term and 
the repeated indices $i,j,k$ etc.\ on the right-hand side are summed over.
Various quantities are defined as follows:\cite{Gao}
\begin{equation}
\begin{split}
{\bm P}_{\rm E} &= \frac{1}{4} \left[ \int ({\bm B}\times {\hat{\bm D}}) \Uell^\dagger
( {\hat {\bm V}} + {\bm v}_0 )\cdot  ({\bm B}\times {\hat{\bm D}}) \Uell d{\bm r} 
+ {\rm c.c.} \right], \cr
G_{\ell', \ell} &= -{\bm B}\cdot {\bm M}_{\ell', \ell}, \qquad
{\bm M}_{\ell', \ell} = \frac{1}{2}(\sum_{\ell''\ne \ell} {\bm V}_{\ell', \ell''} \times {\bm A}_{\ell'', \ell} 
+ {\bm v}_0 \times {\bm A}_{\ell', \ell}), \cr
g_{ii} &= \int \frac{\partial \Uell^\dagger}{\partial k_i} 
\frac{\partial \Uell}{\partial k_j} d{\bm r}, \qquad 
\alpha_{ij} = \frac{\partial^2 \Eell}{\partial k_i \partial k_j},
\label{GaoAll}
\end{split}
\end{equation}
where ${\hat {\bm V}} =-\frac{i}{\hbar} [  {\hat {\bm r}}, {\hat H}_0 ], 
{\bm v}_0 = \int \Uell^\dagger {\hat {\bm V}} \ \Uell d{\bm r}, 
{\bm V}_{\ell,\ell'} = \int \Uell^\dagger {\hat {\bm V}} \ \Uellp d{\bm r}$, 
and ${\hat{\bm D}} = \frac{\partial}{\partial {\bm k}} + i{\bm a}_0$. 
The quantities, ${\bm a}_0$ and ${\bm A}_{\ell,\ell'}$ are intraband and interband 
Berry connection, respectively, defined as
\begin{equation}
{\bm a}_0 = i \int \Uell^\dagger  \frac{\partial \Uell}{\partial {\bm k}} d{\bm r}, \quad
{\bm A}_{\ell,\ell'} = i \int \Uell^\dagger  \frac{\partial \Uellp}{\partial {\bm k}} d{\bm r}.
\end{equation}


In our Hamiltonian with $V(-{\bm r})=V({\bm r})$ and with time-reversal symmetry, 
${\bm a}_0$ is equal to zero as shown in (\ref{AppBn2}). 
${\bm A}_{\ell,\ell'}$ is a well-known quantity in the literatures,\cite{Blount2,WilsonText,HS2}
and it appears in $p_{\ell\ell', \mu}$ (eq.~(\ref{Joffdiagonal})) in the present formalism. 
Gao {\it et al} defined ${\hat H}_0$ as the Hamiltonian without external field.\cite{Gao} 
In the present notation, it should be $H_{\bm k}$. 
Then, the velocity operator $\hat {\bm V}$ is just $\partial H_{\bm k}/\partial {\bm k}$. 
Therefore, their ${\bm v}_0$ should be simply equal to $\partial \Eell/\partial {\bm k}$ since
\begin{equation}
{\bm v}_0 = \int \Uell^\dagger \frac{\partial H_{\bm k}}{\partial {\bm k}} \Uell d{\bm r} 
=\frac{\partial \Eell}{\partial {\bm k}},
\end{equation}
as shown in (\ref{Jmatrixelement}). 
Then, the $\mu$ component of ${\bm P}_{\rm E}$ can be rewritten as 
\begin{equation}
\begin{split}
{P}_{{\rm E}, \mu} &= \frac{1}{4} \biggl[ \varepsilon_{\mu\nu\tau} \varepsilon_{\mu'\nu'\tau'} 
B_\nu B_{\nu'} 
\int \frac{\partial \Uell^\dagger}{\partial k_\tau} 
\left( \frac{\partial H_{\bm k}}{\partial k_{\mu'}} + \frac{\partial \Eell}{\partial k_{\mu'}} 
\right) \frac{\partial \Uell}{\partial k_{\tau'}} d{\bm r} + {\rm c.c.} \biggr].
\end{split}
\end{equation}
By putting ${\bm B}=(0,0,B)$, 
we can show that the second term in (\ref{GaoChi}) corresponds to $\chi_{\rm FS}$, 
but its coefficient is half of our result. 

Next, let us study ${\bm M}_{\ell', \ell}$. 
We can rewrite the $\mu$ component of ${\bm M}_{\ell', \ell}$ as
\begin{equation}
\begin{split}
M_{\ell', \ell\mu} &= \frac{i}{2} \varepsilon_{\mu\nu\tau}
\biggl( \sum_{\ell''\ne \ell} \int \Uellp^\dagger \frac{\partial H_{\bm k}}{\partial k_\nu} \Uellpp
\int \Uellpp^\dagger  \frac{\partial \Uell}{\partial k_\tau} d{\bm r} 
+ \frac{\partial \Eell}{\partial k_\nu} 
\int \Uellp^\dagger  \frac{\partial \Uell}{\partial k_\tau} d{\bm r} \biggr).
\end{split}
\end{equation}
By using the completeness property of $\Uellpp$,  
we can show that $M_{\ell', \ell\mu}$ is rewritten as 
\begin{equation}
\begin{split}
M_{\ell', \ell\mu} &= \frac{i}{2} \varepsilon_{\mu\nu\tau} 
\int \Uellp^\dagger \left( \frac{\partial H_{\bm k}}{\partial k_\nu} + 
\frac{\partial \Eell}{\partial k_\nu} \right) \frac{\partial \Uell}{\partial k_\tau} d{\bm r},
\end{split}
\end{equation}
in our case. 
Therefore, the third term in (\ref{GaoChi}) turns out to be exactly equal to $\chi_{\rm inter}$.  

Similarly, the total of fourth and fifth terms in (\ref{GaoChi}) is equal to $\chi_{\rm occ}$ in our results. 
Thus we can show that Gao {\it et al}'s results (\ref{GaoChi}) are almost equivalent to our result 
except for the coefficient of $\chi_{\rm FS}$.  

\twocolumn

\def\journal#1#2#3#4{#1 {\bf #2}, #3 (#4)}
\def\PR{Phys.\ Rev.}
\def\PRB{Phys.\ Rev.\ B}
\def\PRL{Phys.\ Rev.\ Lett.}
\def\JPSJ{J.\ Phys.\ Soc.\ Jpn.}
\def\PTP{Prog.\ Theor.\ Phys.}
\def\JPCS{J.\ Phys.\ Chem.\ Solids}

\end{document}